\documentclass[prb,twocolumn,groupedaddress]{revtex4-1}

\usepackage{amsmath}
\usepackage{amssymb}
\usepackage{amsfonts}
\usepackage{graphicx}
\usepackage{dcolumn}
\usepackage{float}
\usepackage{bm}
\usepackage{mathrsfs}
\usepackage{txfonts}
\usepackage{tabularx}
\usepackage{bigstrut}
\usepackage[usenames,dvipsnames]{color}
\usepackage[hyperindex,pdftex]{hyperref}
\newcommand{\beq}{\begin{eqnarray}}
\newcommand{\eeq}{\end{eqnarray}}
\newcommand{\ua}{\uparrow}
\newcommand{\da}{\downarrow}
\newcommand{\dg}{\dagger}
\newcommand{\mbf}{\mathbf}
\newcommand{\la}{\langle}
\newcommand{\ra}{\rangle}
\newcommand{\bso}{\boldsymbol}

\begin{document}

\title{The role of impurities on the optical properties of rectangular graphene flakes}

\author{Z. S. Sadeq }
\email{sadeqz@physics.utoronto.ca}
\author{Rodrigo A. Muniz}
\author{J. E. Sipe}

%\thanks{}

\affiliation{Department of Physics, University of Toronto, Toronto, Ontario M5S 1A7, Canada}

\date{\today}

\begin{abstract}
%Graphene flakes (GFs) are an inevitable byproduct of modern synthesis methods for graphene. GFs and other graphene nanostructures, such as nanotubes and nanoribbons, have garnered a lot of interest because of finite size effects, which lead to electronic and optical properties that differ dramatically from bulk graphene. 

%The electronic and optical properties of graphene flakes (GFs) cannot be described by mean field theory alone. 

We study rectangular graphene flakes using mean field states as the basis for a configuration interaction calculation, which allows us to analyze the low lying electronic excited states including electron correlations beyond the mean field level. 
We find that the lowest energy transition is polarized along the long axis of the flake, but the charge distributions involved in these transitions are invariably localized on the zig-zag edges. 
%Typical synthesis methods for graphene flakes can often introduce a variety of impurities on the flake. 
We also investigate the impact of both short and long range impurity potentials on the optical properties of these systems. We predict that even a weak impurity localized at a zig-zag edge of the flake can have a significant -- and often dramatic -- effect on its optical properties. This is in contrast to impurities localized at armchair edges or central regions of the flake, for which we predict almost no change to the optical properties of the flake even with strong impurity potentials. 
\end{abstract}

%\pacs{31.15.aq, 31.15.vq, 31.15.xm}

\maketitle

\section{Introduction}

Graphene has attracted tremendous interest due to its remarkable properties, such as its optical response, its mechanical strength, its zero band-gap, and its thermal conductivity \cite{K.S.Novoselov2004, K.S.Novoselov2005,Y.Zhang2005, A.Rycerz2007,C.Lee2008,D.A.Dikin2007,G.Xin2015, K.F.Mak2008,Y.Zhang2009}. Synthesis methods for graphene often result in small, finite sized byproducts, known as graphene flakes (GFs) or graphene quantum dots, which can be smaller than $2 \,$nm in diameter \cite{M.Buzaglo2016}.  
Some of these flakes have been synthesized and characterized in solution \cite{I.Ozfidan2014,H.Riesen2014,C.Sun2015}, while others have been deposited on substrates such as silicon carbide \cite{L.Liao2010,Y.M.Lin2010}. 
These finite sized flakes and other carbon based materials, with their non-zero energy gaps, have been discussed for possible electronic and optical device applications \cite{M.B.Smith2013,I.Paci2006}. 
Key to the development of GFs for such applications is an understanding of their optical properties. While the optical properties of graphene have been extensively studied \cite{A.H.CastroNeto2009,C.Salazar2016,J.L.Cheng2014, J.L.Cheng2015, J.L.Cheng2017} including some finite size\cite{Y.Li2015,S.Thongrattanasiri2012,Z.Z.Zhang2008} and impurity \cite{RAMuniz2010,HPDahal2011,YCChang2011} effects , the optical properties of finite size graphene flakes have not been, partly due to the shortcomings of mean field theory in finite systems. Previous studies \cite{I.Ozfidan2014,P.Potasz2012,I.Ozfidan2015, I.Ozfidan2016,I.Ozfidan2016a,Y.Lu2013,Y.Lu2016, Y.Lu2016a,C.M.Wettstein2016,A.Altintas2017} have demonstrated that the size, shape, and the nature of the edges have an impact on the optical properties of these GFs. 
Earlier work \cite{P.Potasz2012,C.M.Wettstein2016, A.Altintas2017} has focused on hexagonal and triangular flakes with either zig-zag or armchair edges only. 
Rectangular flakes inevitably contain both types of edges, and can display unique behavior due to the competition between effects associated with each kind of edge. 
Research on these flakes has so far been at the mean field level \cite{J.A.Verges2015,J.A.Verges2016,Y.Lu2016a}, and there is still little understanding of the optical properties of these flakes as the size of zig-zag or armchair edges are increased.

%Comprehensive studies of the spatial distribution of electrons involved in the optical transitions, specifically whether they involve charge concentration on the zig-zag or armchair edge of these GFs, have not been undertaken. 

Recently, it has been shown that common methods for generating graphene and GFs can introduce a variety of localized impurities  \cite{C.H.A.Wong2014} in the sample, and these impurities can have a significant effect on the electronic and optical properties. While there has been some work done on how long-range disorder affects the absorption spectra of large armchair edge hexagonal GFs \cite{A.Altintas2017}, there has been little discussion of how localized impurities affect the electronic properties of GFs in general. 

In this paper, we use an extended Hubbard model, also known as the Pariser-Parr-Pople (PPP) model \cite{R.Pariser1953,R.Pariser1953a,Pople1954}, to describe the $p_{z}$ electrons in these GFs, and apply the configuration interaction (CI) method to solve for the many-body states in these systems. We verify that including electron correlations beyond mean-field theory is essential.
We show that varying the size of armchair or zig-zag edges significantly changes the optical properties of these flakes, as well as the nature of the electron distribution involved in the optical transitions. 
We also investigate the effect of impurity potentials of various strengths and ranges in these systems, and demonstrate that impurity potentials located on the zig-zag edges can have a significant effect on the low energy absorption spectrum of these flakes. This is in contrast to impurities located near the center or on armchair edges of the flakes, which have an almost negligible impact on the absorption spectrum regardless of the strength and range of the impurity potential. 

This paper is organized as follows, in Sec. \ref{sec-mod} we present details on the model we have used to solve for the many-body states of GFs, in Sec. \ref{sec-res} we calculate the absorption spectra of two different families of rectangular graphene flakes, in Sec. \ref{sec-imp} we detail the effects of impurities centered at various locations of the GF, and in Sec. \ref{sec-conc} we present our conclusions.  

\section{\label{sec-mod} Method}

%\subsection{Model} 

In graphene and other conjugated organic systems, the $p_{z}$ electrons on the carbon backbone are primarily responsible for the low energy physics, while the $s$, $p_{x}$, $p_{y}$ electrons are primarily responsible for the mechanical stability of the system \cite{Brown2011,Platt1964,Agranovich2009}. We model the $p_{z}$ electrons in the GFs using the Pariser-Parr-Pople (PPP) Hamiltonian \cite{C.Raghu2002, C.Raghu2002a,Barford2005,R.Pariser1953,R.Pariser1953a,Pople1954}:
\beq
H = H_{TB} + H_{Hu} + H_{ext} + H_{imp},
\label{totham}
\eeq
where $H_{TB}$ is the tight-binding Hamiltonian, $H_{Hu}$ is the Hubbard Hamiltonian, $H_{ext}$ extends the Hubbard Hamiltonian, and $H_{imp}$ is the impurity Hamiltonian,
\begin{align} 
 H_{TB} = & - t \sum_{\la i,j \ra, \sigma} c^{\dg}_{i\sigma} c_{j\sigma} \label{TBeq}, \\
H_{Hu} = & U \sum_{i} n_{i\ua} n_{i\da} \label{hubb}, \\
H_{ext} = & \frac{1}{2} \sum_{\substack{i \neq j \\ \sigma \sigma'}} V_{ij} \left( n_{i\sigma} - \frac{1}{2} \right) \left(n_{j\sigma'} - \frac{1}{2} \right), \label{exthubb} \\
H_{imp} = & \sum_{i\sigma} \overline{\varepsilon}_{i}\left(\mbf{r}_{c} \right) c^{\dg}_{i\sigma} c_{i\sigma}. \label{impham}
\end{align} 
Here $t = 2.66 \,e$V is the hopping parameter \cite{Kundu2009,J.Hachmann2007}, $\sigma$ is a spin label, $i$ and $j$ are site labels, and the angular brackets indicate sums over nearest neighbors only. 
The fermion creation and annihilation operators are denoted respectively by $c^{\dg}_{i\sigma}$ and $c_{i\sigma}$, so the electron number operator for spin $\sigma$ and site $i$ is $n_{i\sigma} = c^{\dg}_{i\sigma} c_{i\sigma}$. 

We set the on-site repulsion parameter to $U = 8.29 \,e$V for all calculations \cite{J.A.Verges2015}. While some researchers \cite{Y.Lu2013,Y.Lu2016,Y.Lu2016a,A.Altintas2017} have used heavily screened values for $U$, we base our choice of this parameter on recent calculations of the Coulomb repulsion parameter in graphene \cite{J.A.Verges2015, T.O.Wehling2011,M.Schuler2013}. %There is no consensus on the value of $U$ used to model the electrons in GFs. 
The parameters we use in this article have been shown to result in a semi-metal solution for the ground state of graphene-like systems \cite{W.Wu2014,S.Arya2015,L.M.Martelo1997}, 
and are also similar to values used in calculations for other organic systems \cite{Barford2005}. 
We approximate the long-range Coulomb repulsion by the Ohno interpolation \cite{Barford2005},
\beq
V_{ij} = \frac{U}{\sqrt{1 + \left( 4\pi \epsilon_{0} U  \epsilon r_{ij} / e^2 \right)^{2}}},
\eeq
where $U$ is the on-site repulsion parameter, $\epsilon$ is a screening parameter, $r_{ij}$ is the distance between sites $i$ and $j$, $e = -|e|$  is the electronic charge, and $\epsilon_{0}$ is the vacuum permittivity. We set $\epsilon = 5$ for all calculations, in accordance with other researchers who have utilized this value of the screening parameter to model the long range Coulomb repulsion in similar systems \cite{J.A.Verges2015,I.Ozfidan2014}. 

We model the potential at $\mbf{r}_{i}$ due to an impurity at $\mbf{r}_{c}$ by a Gaussian
\beq
\overline{\varepsilon}_{i} = \overline{\varepsilon}_{max} \exp \left(- \frac{\left( \mbf{r}_{i} - \mbf{r}_{c} \right)^{2}}{2\tau^{2}} \right),
\eeq
where $\tau$ characterizes the range of the impurity potential, $\mbf{r}_{i}$ is the location of site $i$, and $\overline{\varepsilon}_{max}$ is the maximum value of the impurity potential. While a range of these parameters are considered below, the default parameters we use to model the impurity potential are $\overline{\varepsilon}_{max} = t/3$ and $\tau = l_{b}$. These parameters are in line with recent work done on modeling disorder in large GFs \cite{A.Altintas2017}.

%For a ``medium'' impurity we set $\overline{\varepsilon}_{max} = t/3$, and for a ``weak'' impurity we set $\overline{\varepsilon}_{max} = t/5$. 
%For ``short range'' impurities we set $\tau = l_{b}/5.0$ where $l_{b}$ is the bond length which is set to $l_{b} = 1.42 \AA$, and for ``long range'' impurities we set $\tau = l_{b}$. Unless otherwise specified, we use the parameters $\overline{\varepsilon}_{max} = t/3$ and $\tau = l_{b}$ to model the impurity potential. 

We first solve the Hartree-Fock (HF) equations for the PPP Hamiltonian (\ref{totham}),
\begin{widetext}
\begin{align}
H^{HF} = & -t \sum_{\la i,j \ra, \sigma} c^{\dg}_{i\sigma} c_{j\sigma} + \sum_{i\sigma} \overline{\varepsilon}_{i}c^{\dg}_{i\sigma} c_{i\sigma}  \nonumber 
\\
& + U \sum_{i}  \la n_{i\ua} \ra n_{i\da} + \la n_{i\da} \ra n_{i\ua} - \la n_{i\ua} \ra \la n_{i\da} \ra - \la c^{\dg}_{i\ua} c_{i\da} \ra c^{\dg}_{i\da} c_{i\ua} - \la c^{\dg}_{i\da} c_{i\ua} \ra  c^{\dg}_{i\ua} c_{i\da} + \la  c^{\dg}_{i\ua} c_{i\da} \ra \la  c^{\dg}_{i\da} c_{i\ua} \ra \nonumber 
\\
& + \sum_{i \neq j} V_{ij} \left( n_{i} \la n_{j} \ra  - n_{i} 
- \frac{1}{2} \la n_{i} \ra \la n_{j} \ra + \frac{1}{2}
- \frac{1}{2} \sum_{\sigma\sigma'} \la c^{\dg}_{i\sigma} c_{j\sigma'} \ra c^{\dg}_{j\sigma'} c_{i\sigma} + \la c^{\dg}_{j\sigma'} c_{i\sigma} \ra c^{\dg}_{i\sigma} c_{j\sigma'} - \la c^{\dg}_{i\sigma} c_{j\sigma'} \ra \la c^{\dg}_{j\sigma'} c_{i\sigma}  \ra \right). 
\label{HFeqs}
\end{align}
\end{widetext}
This equation is derived following the prescription found in standard references \cite{J.A.Verges2015,H.Bruus2004,G.F.Giuliani2005,Pines1977}. We diagonalize (\ref{HFeqs}) self consistently, using the tight-binding (\ref{TBeq}) eigenfunctions as an initial guess. For the parameters used, we find two stable solutions: one antiferromagnetic and one paramagnetic. The antiferromagnetic solution is discarded \cite{S.Sorella1992,S.Arya2015,J-P.Malrieu2016,L.M.Martelo1997,F.Plasser2013}, since other methods that treat electron correlation more rigorously than HF, such as Quantum Monte Carlo and CI calculations, have shown that the paramagnetic solution \cite{F.Moscardo2009,E.San-Fabian2011,W.Wu2014} is the lower energy state on similar but smaller systems. 
%The antiferromagnetic solution being lower in energy is an artifact of HF. 
%In general, solutions to the HF equations (\ref{HFeqs}) can yield states that are superpositions of spin up and spin down states; for the systems of interest in this paper, the solutions to the HF equations are spin separated, i.e. states that are either spin up or spin down.
Upon the self-consistent solution of (\ref{HFeqs}) with paramagnetic expectation values, one can write (\ref{HFeqs}) in its diagonal form
\beq
H^{HF} = \sum_{m \sigma} \hbar \omega_{m\sigma} C^{\dg}_{m\sigma} C_{m\sigma},
\eeq
where $\hbar \omega_{m\sigma}$ are the eigenvalues associated with the single particle states. The operators $C^{\dg}_{m\sigma}$ and $C_{m\sigma}$ can be written in terms of the site basis as
\beq
&& C^{\dg}_{m\sigma} = \sum_{i} M_{m\sigma,i} c^{\dg}_{i\sigma}, \label{eq-sphf}\\
&& C_{m\sigma} = \sum_{i} M^{*}_{m\sigma,i} c_{i\sigma}, \label{eq-sphf2}
\eeq
where $C^{\dg}_{m\sigma}$ indicates the creation of a HF quasiparticle in state $m$ with spin $\sigma$, $M_{m\sigma,i}$ is the amplitude associated with the state $m$ at site $i$, and is typically non-zero for all $i$. 
% The HF quasiparticle operators obey the fermionic anticommutation relations $\{ C_{i\sigma}, C^{\dg}_{j\sigma'} \} = \delta_{ij}\delta_{\sigma\sigma'}$.

The single particle states obtained from solving the HF equations with paramagnetic expectation values self consistently are then used to construct the HF ground state
\beq
|g_{HF} \ra = \prod^{N/2}_{m} C^{\dg}_{m\ua}  C^{\dg}_{m\da} |\text{vac} \ra,
\label{HFgs}
\eeq
where $|\text{vac} \ra$ represents the full vacuum, and $N$ is the number of electrons in the system. 
The states that are filled in the HF ground state are denoted as ``valence", and those that are unfilled in the HF ground state are denoted as ``conduction". We denote the highest occupied HF state as HOHF, and the $n^{th}$ state below that the HOHF-n orbital. Similarly, we denote the lowest unoccupied HF state as LUHF, and the $m^{th}$ state above that the LUHF+m orbital. 

We then rewrite the total Hamiltonian (\ref{totham}) in the HF electron-hole basis; its full form is given in Appendix \ref{app-Ham}. 
We use an electron-hole basis, where the HF electron creation is designated by the operator $a^{\dg}_{L_{m}\sigma}$, and the HF hole creation is designated by the operator $b^{\dg}_{H_{n}\sigma}$, so 
\beq
a_{L_{m}\sigma} = C_{L_{m}\sigma}, 
 \quad  \quad
b^{\dg}_{H_{n}\sigma} = C_{H_{n}\tilde{\sigma}}, \label{hfhole}
\eeq
where $\tilde{\sigma}$ is the opposite spin of $\sigma$, $L_{m}$ is the LUHF+m orbital, and $H_{n}$ is the HOHF-n orbital. The LUHF state itself is denoted as $L_{0}$ and the HOHF is denoted as $H_{0}$, or for simplicity, $L$ and $H$ respectively. 

We select an ``active space" for our CI calculation defined by a set of HF excited states, identified by overbars. The singly excited states are of the form
\beq
|\overline{L_{m},H_{n};\sigma}\ra = a^{\dg}_{L_{m}\sigma} b^{\dg}_{H_{n}\tilde{\sigma}}|g_{HF} \ra,
\label{eq-hfsing}
\eeq
where $m$, $n$ range over $\{ 0, \dots ,4 \}$. The doubly excited states are of the form
\beq
|\overline{L_{m}L_{m'};H_{n},H_{n'}}\ra = a^{\dg}_{L_{m}\ua} a^{\dg}_{L_{m}\da}b^{\dg}_{H_{n}\da} b^{\dg}_{H_{n}\ua}|g_{HF} \ra, \label{hfdoubs}
\eeq
where $m,m',n,n'$ all range over $\{ 0, \dots ,4 \}$. In the special case where $m = m'$ and $n = n'$ we write $|\overline{2L_{m} H_{n}}\ra$ for $|\overline{L_{n} L_{n};H_{m} H_{m}}\ra$. Together with the HF ground state (\ref{HFgs}), these HF excited states are used to approximately diagonalize the total Hamiltonian (\ref{totham}).  Upon diagonalization of the many-body Hamiltonian (\ref{totham}), the states become superpositions of the HF states; for example, the ground state is given by
\begin{align}
|g \ra = & f^{GS}_{g}|g_{HF} \ra + \sum_{\alpha,\beta, \sigma} f^{\alpha\beta\sigma} _{g} |\overline{L_{\alpha}, H_{\beta};\sigma}\ra \nonumber \\
& + \sum_{\alpha\beta\gamma\delta} f^{\alpha\beta\gamma\delta}_{g} |\overline{L_{\alpha} L_{\beta}; H_{\alpha} H_{\beta}}\ra,
\label{CI_gr_st}
\end{align}
where $f^{GS}_{g}$, $f^{\alpha\beta\sigma}_{g}$, $f^{\alpha\beta\gamma\delta}_{g}$ are the CI ground state amplitudes of the HF ground state, single excitations, and double excitations respectively.  
The excited states have similar expressions. 
We do not include higher order excitations in our calculation because their energies are large  compared to the single and double excitations that are included, so their contribution to the CI low energy states can be expected to be small; as well, the matrix elements of the Hamiltonian between the HF ground state and the states with higher order excitations vanish. 
For the systems of interest, the main contribution to the CI ground state (\ref{CI_gr_st}) is from the HF ground state, as it corresponds to more than half $\left( \left|f^{GS}_{g} \right|^2 >0.5 \right)$  for all the systems studied in this article, indicating that the HF ground state is indeed a good starting point. We find that increasing the active space in our calculation does not significantly change the composition of the low energy states, indicating that our calculation already includes all relevant HF excitations that make up the low energy states of the system.

%The contribution from the HF ground state to the CI ground state is $|f^{GS}_{g}|^{2}$, and we find that increasing the active space beyond the HF states we report does not change this number significantly, which indicates that the CI ground state in our calculation is well converged in regard to the active space.

%\hl{Want to say something like how if} $f_{gs}$ \hl{is too small or fluctuates when active space changes it indicates that the calculation has not converged and thus we have to either include higher order excitations or move to a different computational technique}
\subsection{Absorption Spectrum Calculation}

In this subsection, we outline the calculation for the first order polarizability of the system, and introduce the spatial profile of the transition, a quantity which is used to characterize the charge distributions of electronic states involved in bright transitions. 

%\subsubsection{The Number Operator in the Electron/Hole Basis}

The number operator for a particular site $i$ is defined as
\beq
n_{i} = \sum_{\sigma} c^{\dg}_{i\sigma} c_{i\sigma}.
\label{eq-dop}
\eeq
In the electron-hole basis, it is written as  
\begin{align}
 n_{i} = & \sum_{mm'\sigma} \Gamma_{mm'\sigma,i} \left(a^{\dg}_{m\sigma} a_{m'\sigma} - b^{\dg}_{m'\sigma} b_{m\sigma}  \right) \nonumber \\
& + \sum_{mm'\sigma} \Gamma_{mm'\sigma,i} \left( a^{\dg}_{m\sigma} b^{\dg}_{m'\tilde{\sigma}} + b_{m\tilde{\sigma}}a_{m'\sigma} \right) + \sum_{m\sigma} \Gamma_{mm\sigma,i},
\end{align}
where we have defined
\beq
\Gamma_{mm'\sigma,i} = M_{m\sigma,i} M^{*}_{m'\sigma,i},
\eeq
and $M_{m\sigma,i}$ is the amplitude of the HF state $m$ with spin $\sigma$ at site $i$.
 
%\subsubsection*{The Transition Dipole Operator and the First Order Polarizability}

The dipole moment operator is approximated as
\beq
\bso{\mu}  = \sum_{i} e\mbf{r}_{i} \left( \sum_{\sigma} c^{\dg}_{i\sigma} c_{i\sigma}-1 \right),
\eeq
where the charge of each nucleus not balanced by the in-plane bonding electrons of the molecule is included, and so the dipole moment operator is independent of origin. Transforming it into the electron-hole basis as defined in the previous subsection, we have
\begin{align}
\bso{\mu}  = & \sum_{m\sigma} \bso{\mu}_{mm\sigma}  - e\sum_{i} \mbf{r}_{i} +  \sum_{mm'\sigma} \bso{\mu}_{mm'\sigma} a^{\dg}_{m\sigma} a_{m'\sigma}  \nonumber \\
& - \sum_{mm'\sigma} \bso{\mu}_{mm'\sigma} b^{\dg}_{m'\sigma} b_{m\sigma} +  \sum_{mm'\sigma} \bso{\mu}_{mm'\sigma} a^{\dg}_{m\sigma} b^{\dg}_{m'\tilde{\sigma}} \nonumber \\
& + \sum_{mm'\sigma} \bso{\mu}_{mm'\sigma} b_{m\tilde{\sigma}} a_{m'\sigma},
\end{align}
where  
\beq
\bso{\mu}_{mm'\sigma} = \sum_{i} e \mbf{r}_{i} M_{m\sigma,i} M^{*}_{m'\sigma,i}.
\eeq
%and we exploited the fact that spin is a good quantum number, and each HF state has a spin label.
%the HF states are spin separated.

We determine the absorption spectrum by calculating the imaginary component of the first order polarizability of the system  \cite{Boyd2008}. Assuming the system is initially in the ground state, the imaginary component of the first order polarizability is given by
\beq
\text{Im} \left( \alpha^{(1)}_{kl}(\omega) \right) = \frac{\gamma}{\epsilon_{0} \hbar} \sum_{m} \frac{ \mu_{gn}^{k}\mu_{ng}^{l}}{ \left(\omega_{ng}-\omega\right)^{2}+\gamma^{2}},
\eeq
where $k,l$ are Cartesian components, $\bso{\mu}_{gn}$ is the matrix element of the dipole moment operator between the ground state and the state $n$, and $\hbar\gamma$ is a frequency broadening, which we set to $\hbar\gamma = 0.01 \,e$V for all calculations, primarily for reasons of presentation \cite{Z.S.Sadeq2015}.

%\subsubsection{The Transition Electron Density}

In order to analyze the charge distributions of electronic states involved in bright transitions, we define
\beq
T_{Y;i}= \la g | n_{i} | Y \ra,
\label{eq-tdop}
\eeq
where $i$ is the site, $g$ is the CI ground state, and $Y$ is a CI excited state. This quantity is related to the matrix element of the dipole moment operator between the ground state and $Y$, the ``transition dipole moment'',
\beq
\la g | \bso{\mu} | Y \ra = \sum_{i} e\mbf{r}_{i} T_{Y;i}.
\eeq
We call $T_{Y;i}$, taken as a function of $i$ for fixed $Y$, the spatial profile of the transition $g \rightarrow Y$. 

%In the next sections, we show plots of $T_{Y;i}$ where we place a circle at the location of each site $i$, the area of each circle indicates the magnitude of $T_{Y;i}$, and the color indicates whether it is positive or negative. 

\section{\label{sec-res} Pristine graphene flakes}

We first investigate the optical properties of the pristine GFs. 
In this section, we present our calculations for the absorption spectra of the two families of pristine GFs, and observe the trends as we increase their size. We also present the spatial profiles of a select few low energy transitions. For the rest of this paper, we shall refer to the HF single particle levels as ``modes'', and we shall refer to a state that results from the CI calculation as a ``state''. In these systems, the HOHF and LUHF primarily have electron density concentrated on the zig-zag edges of the flake, and are labeled as edge modes. Modes that are below (above) the HOHF (LUHF) typically have electron density spread throughout the flake, and are labeled as bulk modes. We label the lowest energy bright excited state the $S_{1}$ state, and the second lowest energy bright excited state the $S_{2}$ state, and so on. 

%\subsection{Systems of Interest}

\begin{figure}[htb!]
\begin{center}
\includegraphics[scale=0.28]{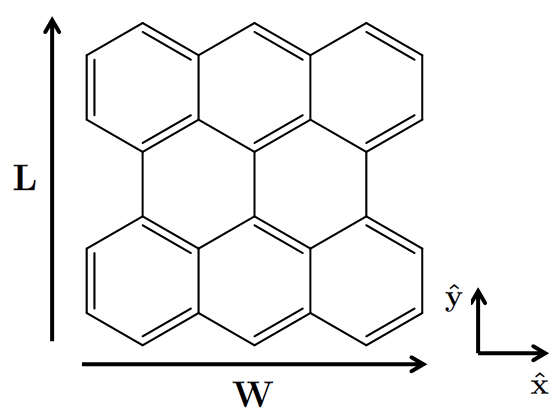}
\caption{A cartoon illustration of the W3L3 flake. Here the width and length of this flake are both composed of three hexagons. The axis convention used in the rest of the paper is also illustrated in this figure; the width corresponds to the $\hat{\mbf{x}}$ axis, and the length corresponds to the $\hat{\mbf{y}}$ axis.} 
\label{figcartoon}
\end{center}
\end{figure}

We investigate rectangular graphene flakes, 
and use two numbers to specify a particular rectangular GF: the ``width'', which identifies  the number of hexagons on the horizontal axis, and the ``length'', which identifies the number of hexagons on the vertical axis of this flake. The notation we use is W$w$L$l$ for a flake with a width of $w$ hexagons, and a length of $l$ hexagons. The width corresponds to the size of the  zig-zag edges, while the length corresponds to the size of the armchair edges of the flakes.  We illustrate the example of the W3L3 flake in Fig. \ref{figcartoon}. 
The four families of flakes we consider are the W3L$n$, W$n$L3, W5L$n$, and W$n$L5 families. The W3L$n$ family consists of all flakes where the width is 3 hexagons but the length varies, e.g. W3L3, W3L5,...,W3L11; in this family of flakes the armchair edges are larger than the zig-zag edges. The W$n$L3 family consists of all flakes where the length is 3 hexagons but the width varies, e.g. W3L3, W5L3,...,W11L3; in this family of flakes the zig-zag edges are larger than the armchair edges. The W5L$n$ family consists of all flakes where the width is 5 hexagons but the length varies, e.g. W5L5, W5L7, and W5L9; in this family of flakes the armchair edges are larger than the zig-zag edges. Finally, the W$n$L5 family consists of all flakes where the length is 5 hexagons but the width varies, e.g. W5L5, W7L5, and W9L5; in this family of flakes the zig-zag edges are larger than the armchair edges.

%The gf3 families consist of all flakes where the shortest dimension has 3 hexagons, e.g. W3L3, W3L5,..., W3L11, as well as W5L3,..., W11L3.  The gf5 family consists of all flakes where the shortest dimension has 5 hexagons, e.g. W5L5, W5L7, and W5L9 as well as W7L5 and W9L5. Within these families there are some flakes that have larger zig-zag edges than armchair edges ($w > l$) and others that have larger armchair edges than zig-zag ($l > w$). 

\begin{figure}[htb!]
\begin{tabular}{c}
\includegraphics[width = 1.0\columnwidth]{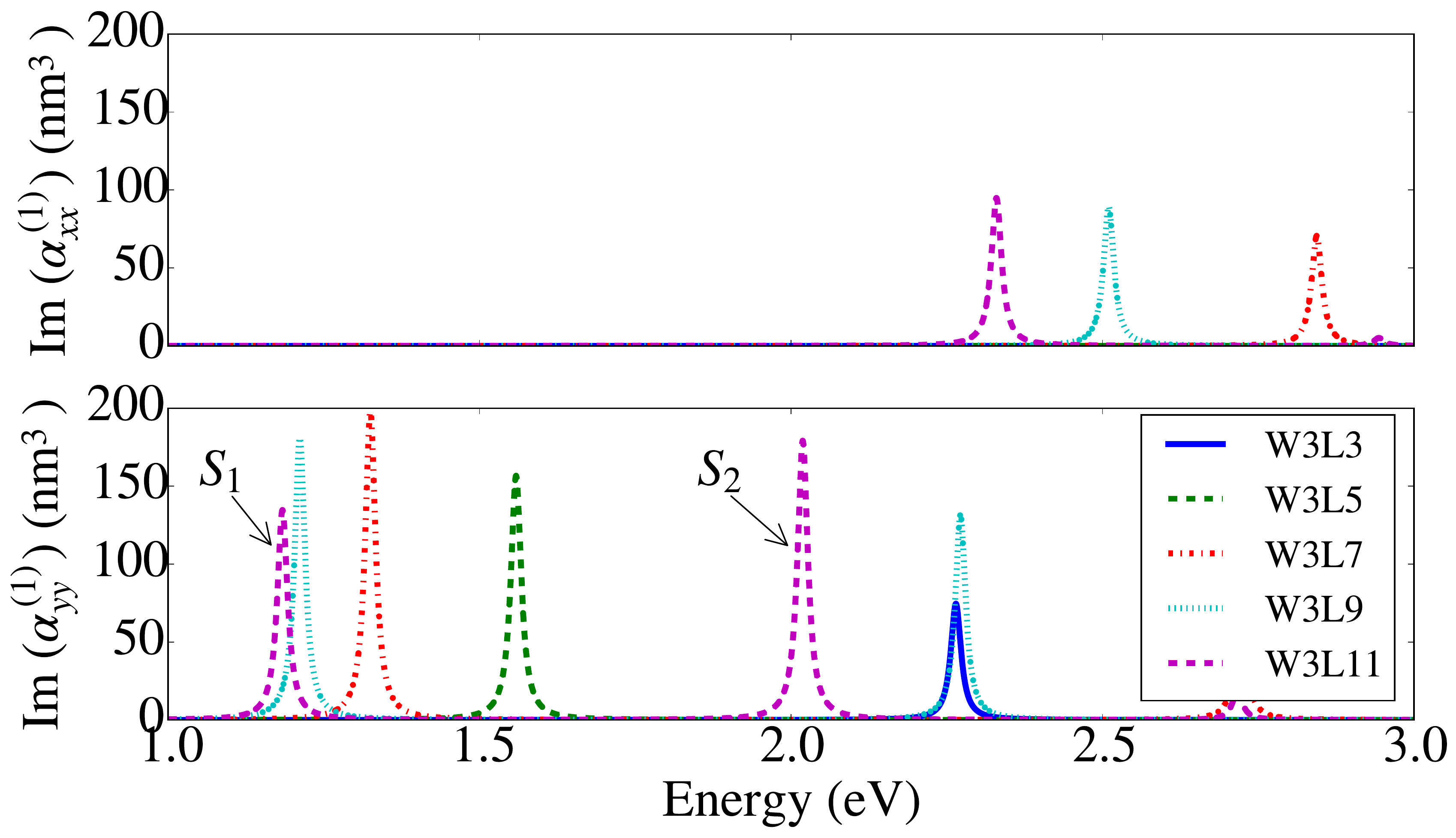} \\
\begin{tabular}{rr}
\includegraphics[width = 0.4\columnwidth]{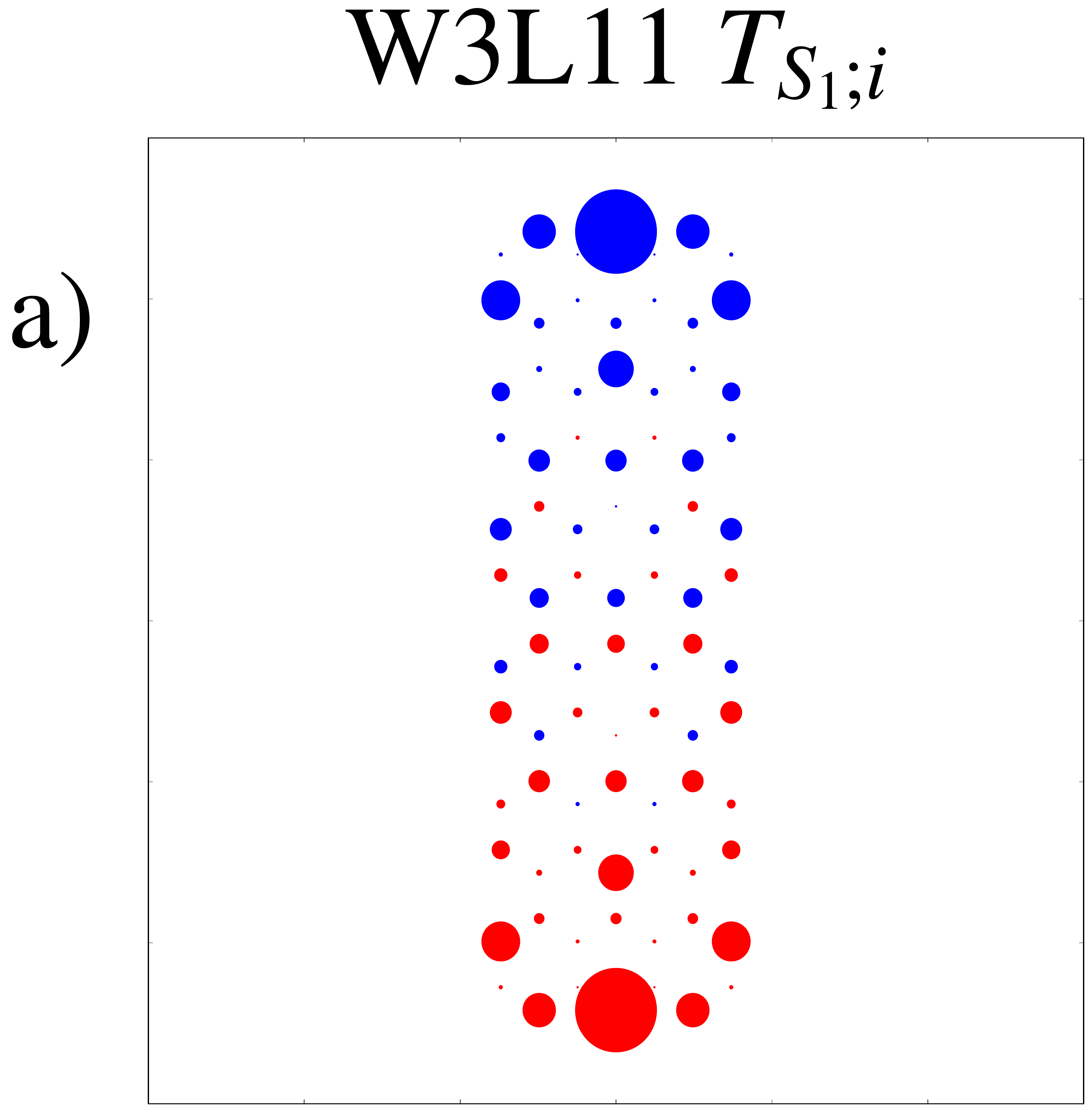} &
\includegraphics[width = 0.4\columnwidth]{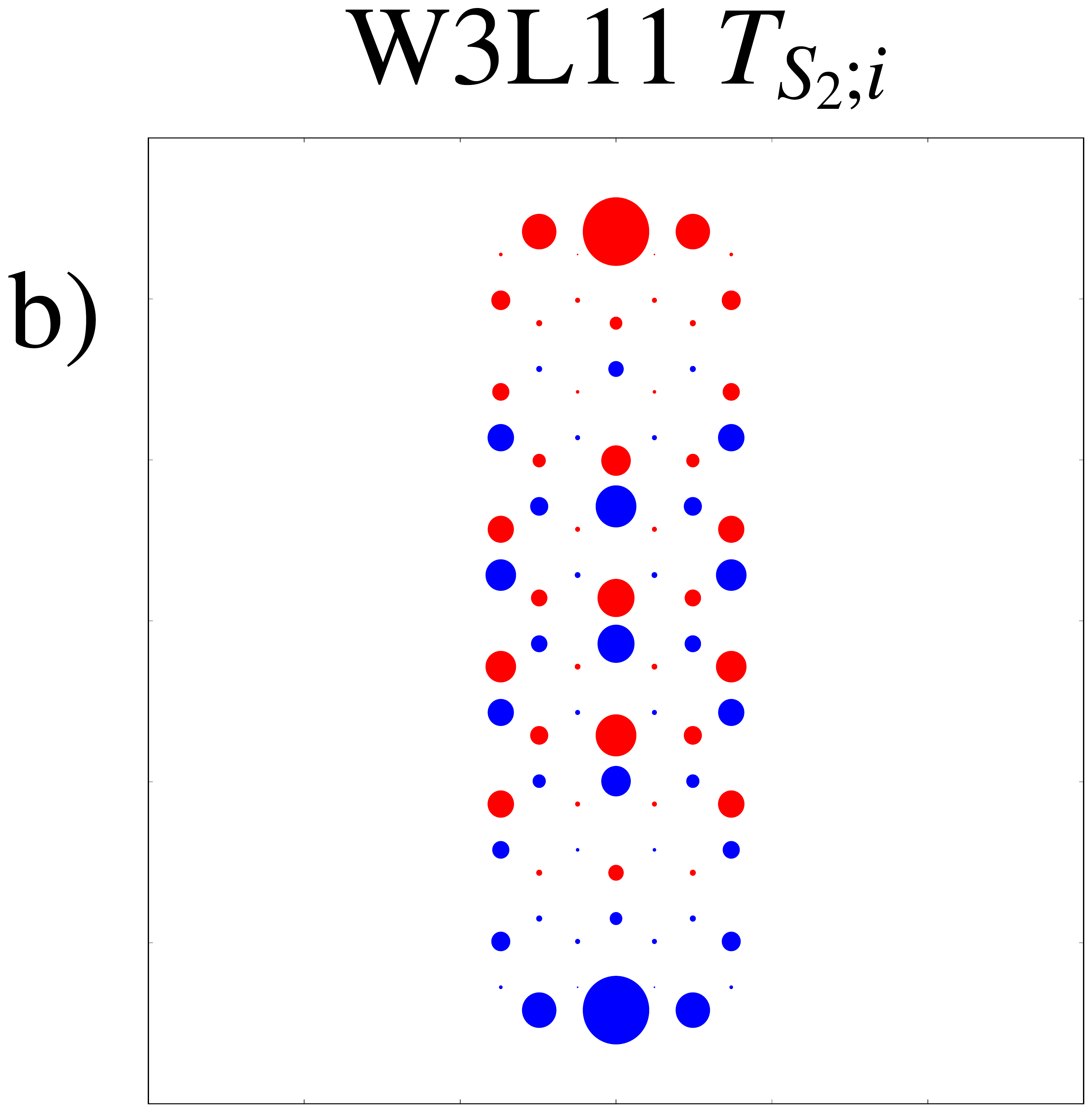}
\end{tabular}
\end{tabular}
\caption{(Top) Absorption spectrum of the W3L$n$ family of flakes. As we consider larger flakes, the first absorption peak, associated with a transition with a transition dipole moment polarized along the long axis (here $\hat{\mbf{y}}$) of the flake, is red shifted. (Bottom) A plot of a) $T_{S_{1};i}$ and b) $T_{S_{2};i}$ in the W3L11 system. For the plots of $T_{S_{1};i}$ and $T_{S_{2};i}$, we place a circle at the location of each site $i$; the area of each circle indicates the magnitude of the relevant quantity, and the color indicates whether it is positive (red) or negative (blue). The majority of the electron concentration is confined to the zig-zag edges for both transitions.}
\label{fig1}
\end{figure}

\begin{figure}[htb!]
\begin{tabular}{c}
\includegraphics[width = 1.0\columnwidth]{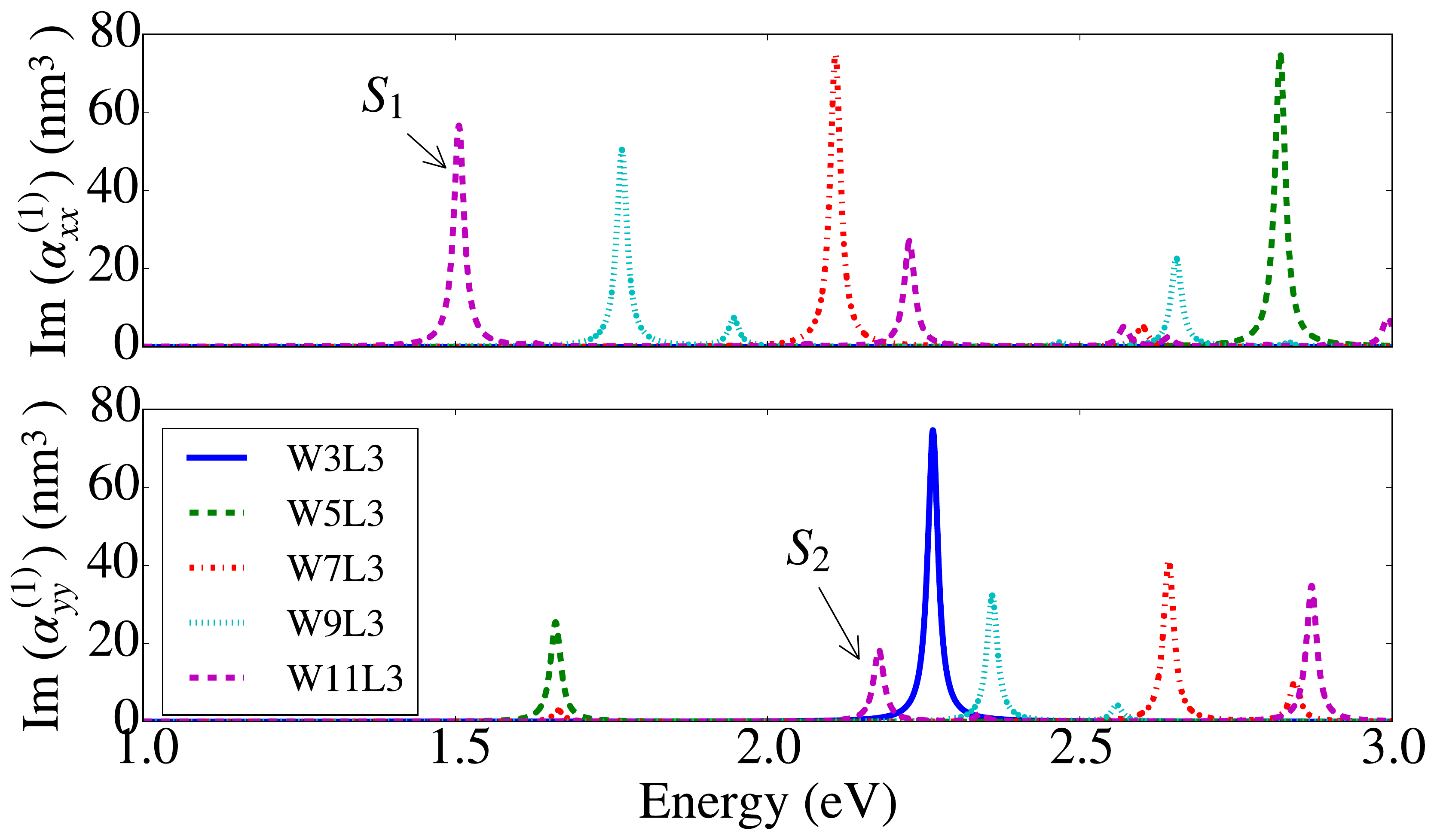} \\
\begin{tabular}{rr}
\includegraphics[width = 0.4\columnwidth]{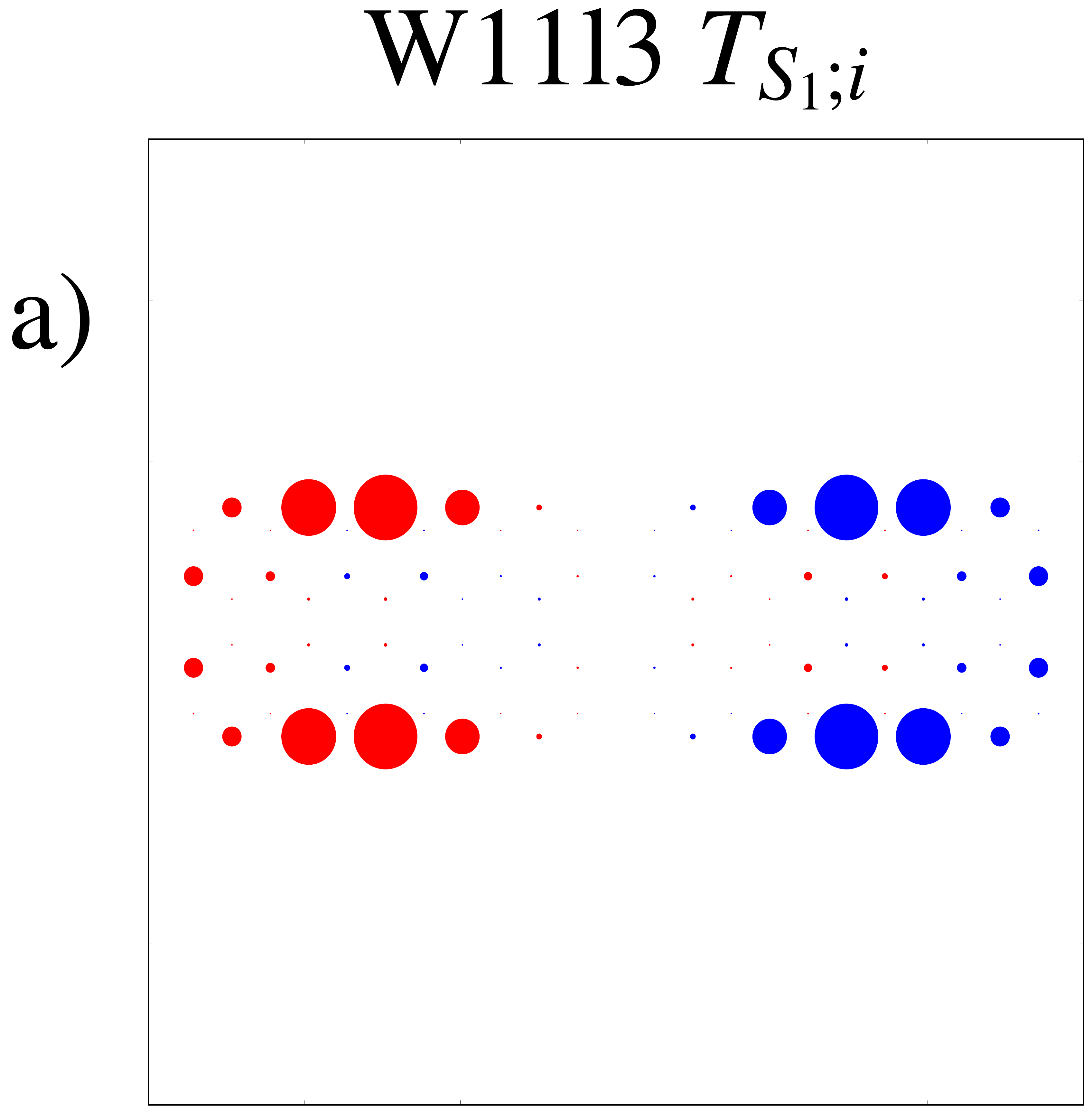} &
\includegraphics[width = 0.4\columnwidth]{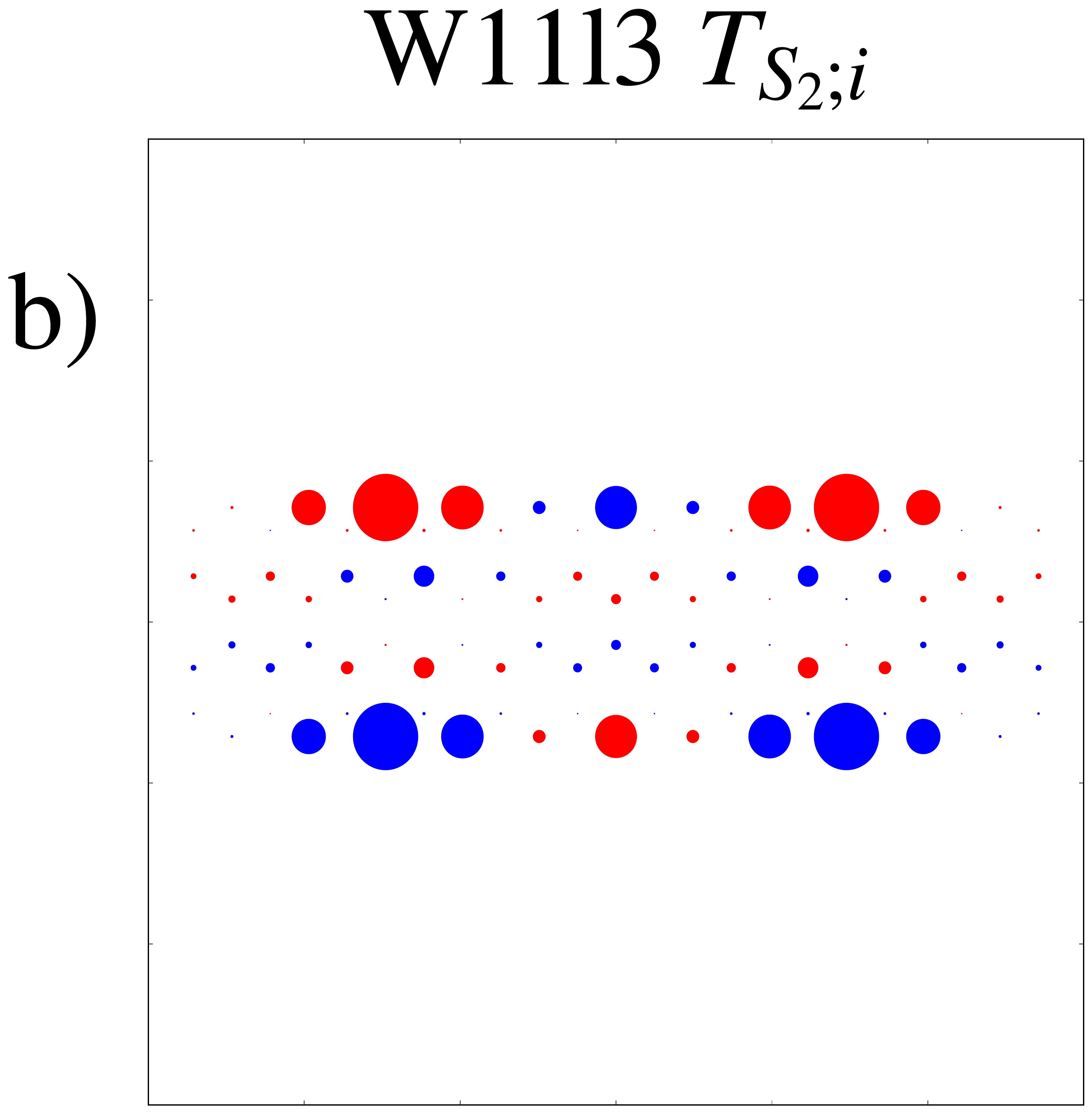}
\end{tabular}
\end{tabular}
\caption{(Top) Absorption spectrum of the W$n$L3 family of flakes. Like the W3L$n$ family, for the large flakes ($n>5$) the first absorption peak is associated with a transition whose transition dipole moment is polarized along the long direction (here $\hat{\mbf{x}}$) of the flake. As the size of the system increases, the first absorption peak is red shifted. (Bottom) A plot of a) $T_{S_{1};i}$ and b) $T_{S_{2};i}$  in the W11L3 system. The majority of the electron concentration is confined to the zig-zag edges for both transitions.}
\label{fig2}
\end{figure}

\begin{figure}[htb!]
\begin{tabular}{c}
\includegraphics[width=1.0\columnwidth]{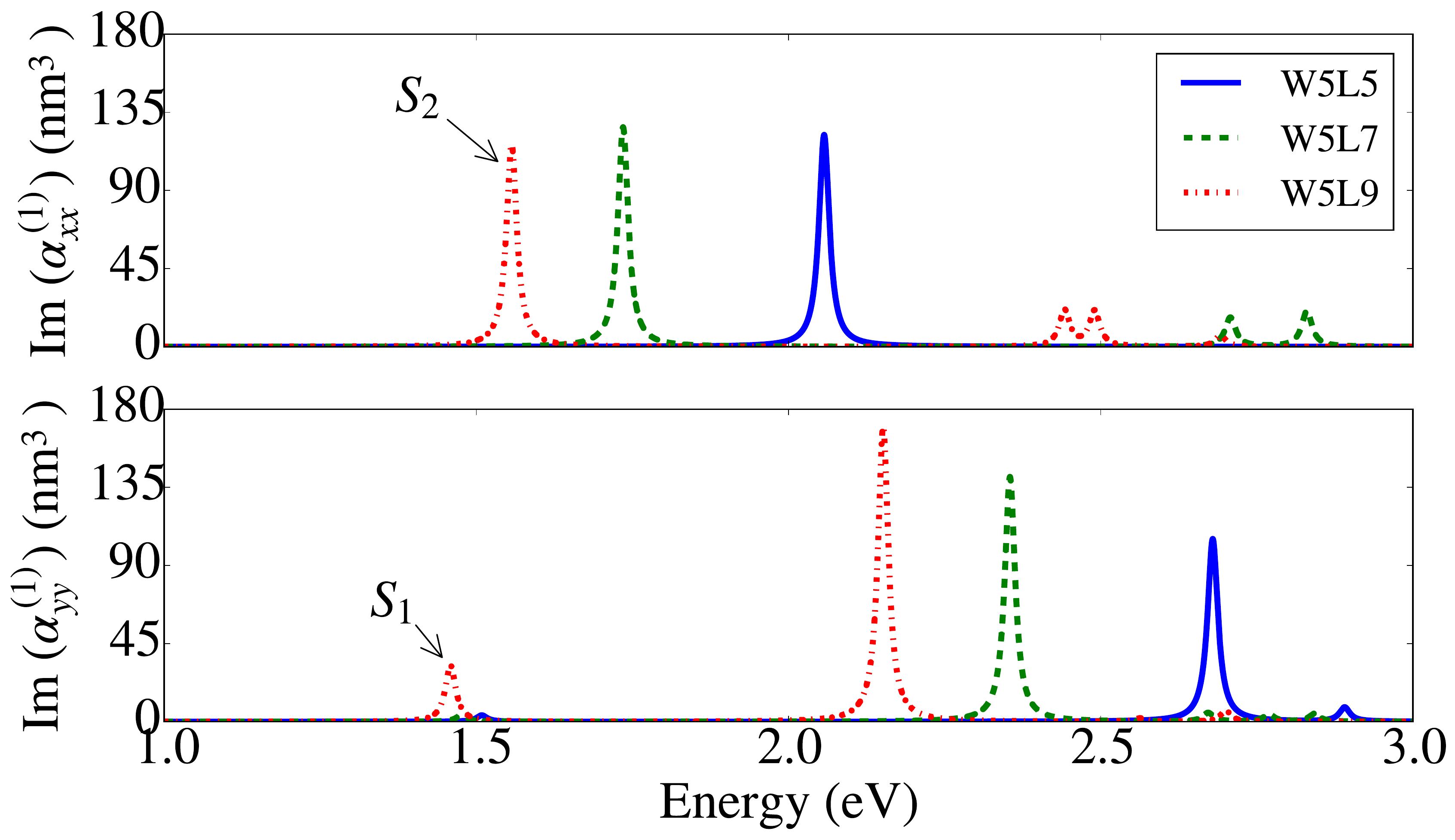} \\
\begin{tabular}{rr}
\includegraphics[width=0.4\columnwidth]{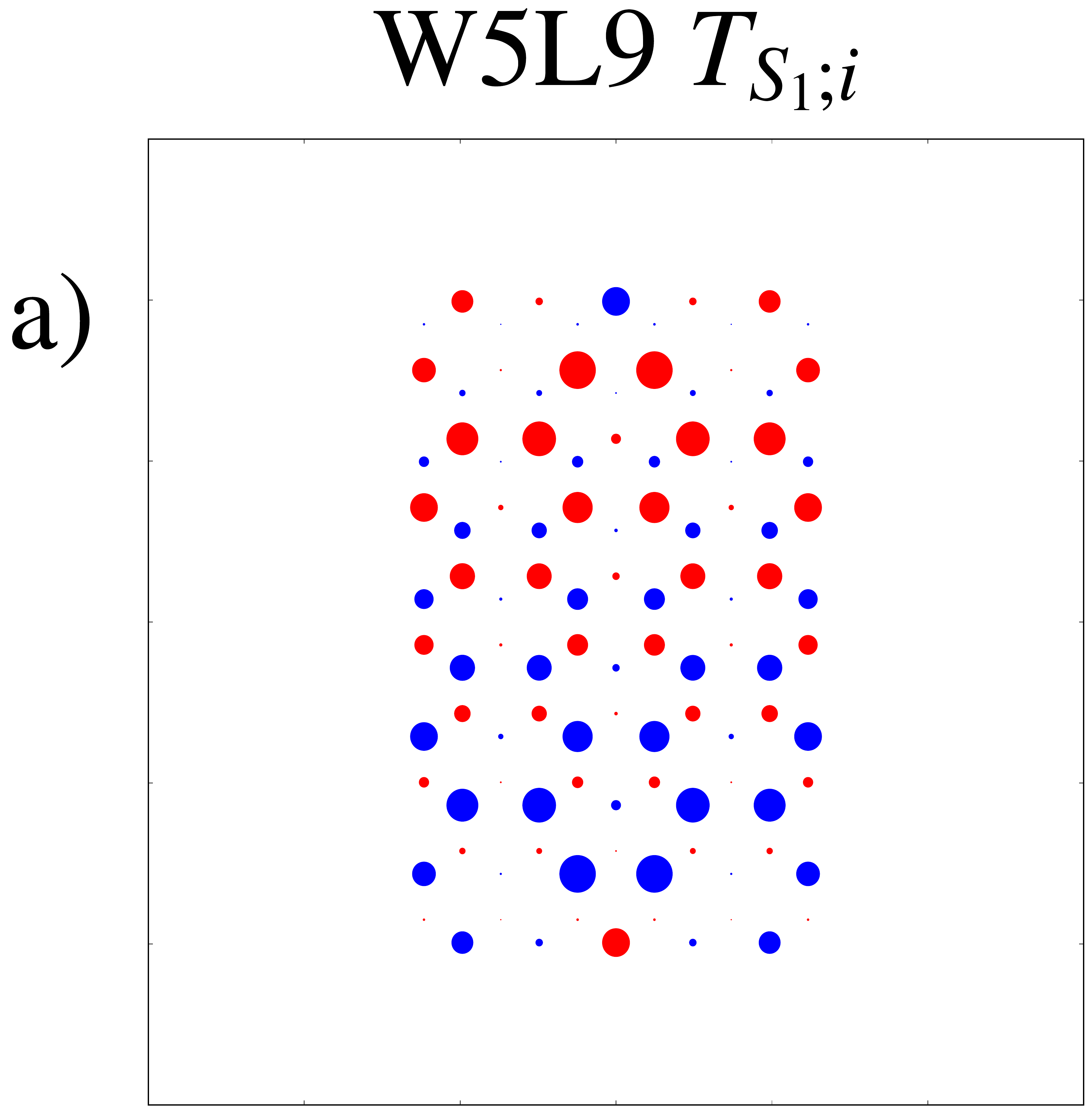} &
\includegraphics[width=0.4\columnwidth]{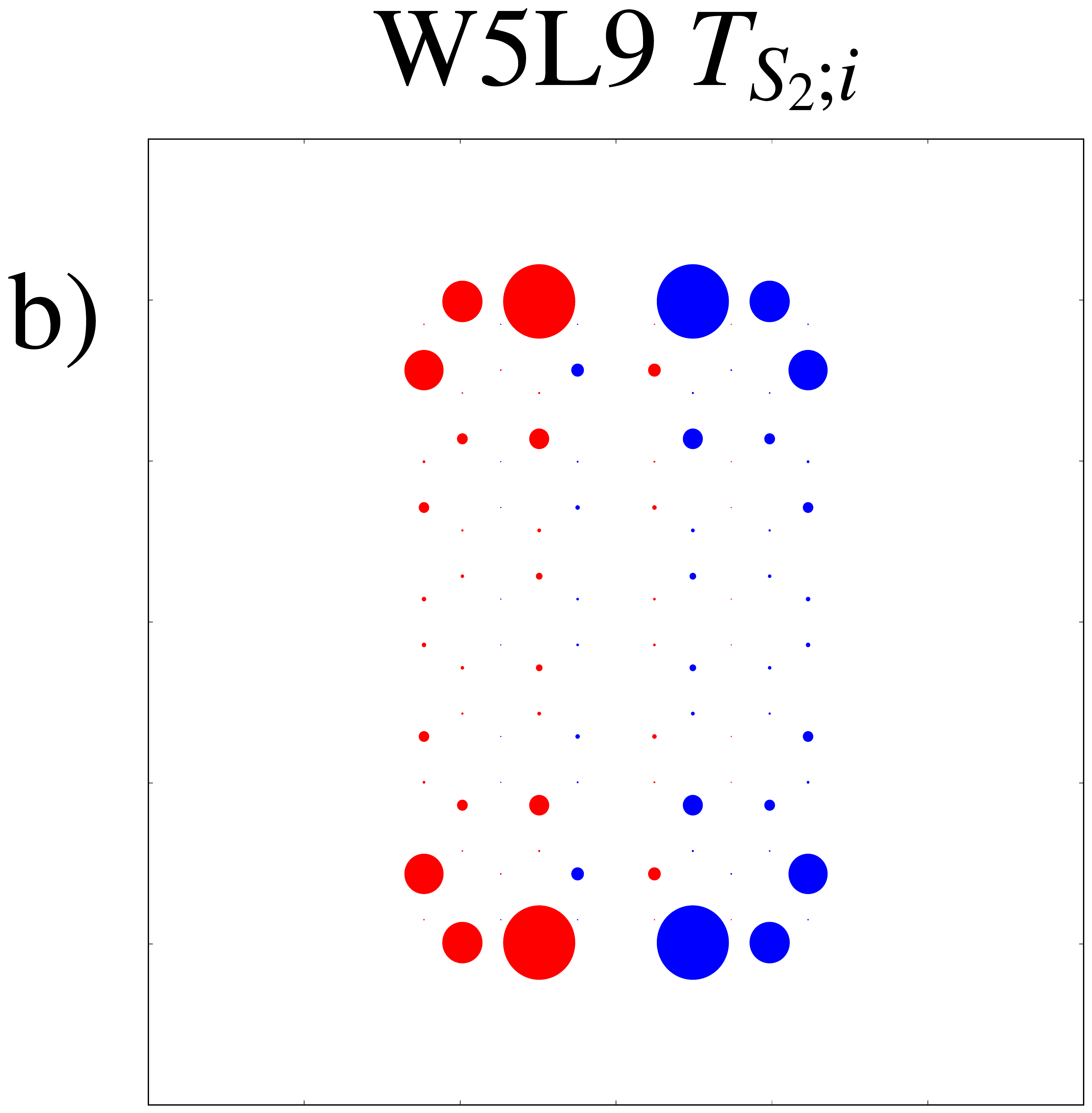}
\end{tabular}
\end{tabular}
\caption{(Top) Absorption spectrum of the W5L$n$ family. As we consider larger flakes, the first two absorption peaks are red shifted. The lowest energy peak is associated with a transition with a transition dipole moment polarized along the long axis (here $\hat{\mbf{y}}$) of the flake, while the second lowest energy peak corresponds to a transition with a transition dipole moment polarized along the short axis (here $\hat{\mbf{x}}$) of the flake. (Bottom) Plot of a) $T_{S_{1};i}$ and b) $T_{S_{2};i}$ for the W5L9 system. The majority of the electron concentration is confined to the zig-zag edges for the brighter transition ($g\rightarrow S_{2}$).}
\label{fig3}
\end{figure}

We first consider the W3L$n$ family of flakes. 
In Fig. \ref{fig1}, we show the absorption spectrum for the W3L$n$ family of flakes. As the size of the flake is increased, the first absorption peak is red shifted, as one would expect even within a non-interacting description of these flakes \cite{A.H.CastroNeto2009,Barford2005}. 
The first absorption peak is associated with a transition with a transition dipole moment that is always polarized along the long (here $\hat{\mbf{y}}$) axis of the flake. 
We show a plot of $T_{S_{1};i}$ for the W3L11 flake in Fig. \ref{fig1}. For this plot, and all subsequent plots of the spatial profiles of the transitions,  we place a circle at the location of each site $i$; the area of each circle indicates the magnitude of $T_{S_{1};i}$, and the color indicates whether it is positive (red) or negative (blue). 
Even though in this family of flakes the size of the armchair edges is larger than the size of the zig-zag edges, $T_{S_{1};i}$ has electron concentration primarily on the zig-zag edges. 
The dominant contributions to the $S_{1}$ state are from single excitations (\ref{eq-hfsing}) of HF quasiparticles from the HOHF to LUHF modes -- two modes that primarily have electron concentration on the zig-zag edges -- with some corrections from excitations of HF quasiparticles from the bulk modes. The second absorption peak is associated with a transition whose transition dipole moment is also polarized along the long axis ($\hat{\mbf{y}}$) of the flake. We show a plot of $T_{S_{2};i}$ for the W3L11 flake in Fig. \ref{fig1}. Similar to $T_{S_{1};i}$, $T_{S_{2};i}$ also has significant electron concentration on the zig-zag edges. The dominant contributions to the $S_{2}$ state are from double excitations (\ref{hfdoubs}) of HF quasiparticles, primarily involving excitations between edge modes.

\begin{figure}[htb!]
\begin{tabular}{c}
\includegraphics[width=1.0\columnwidth]{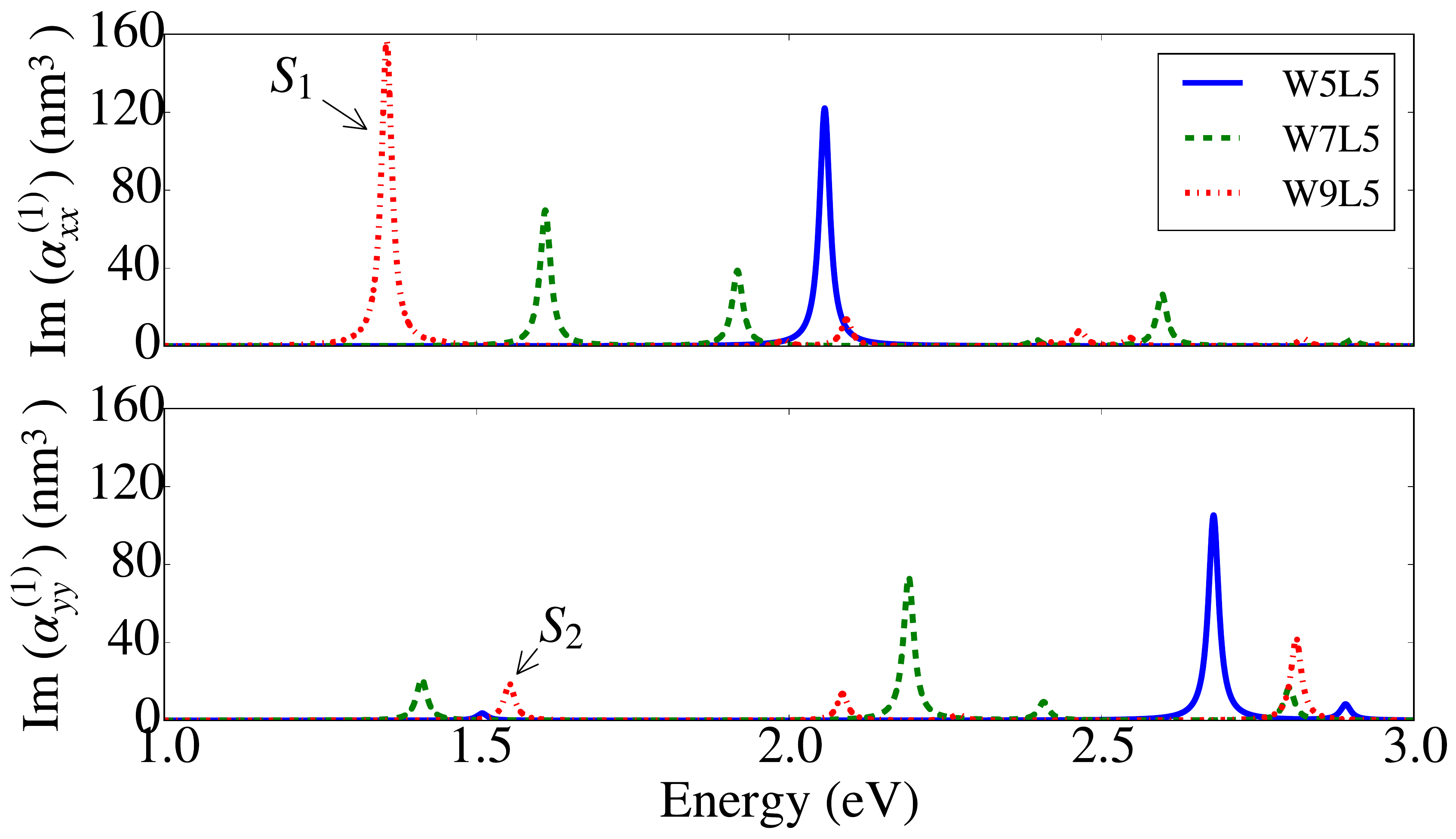} \\
\begin{tabular}{rr}
\includegraphics[width=0.4\columnwidth]{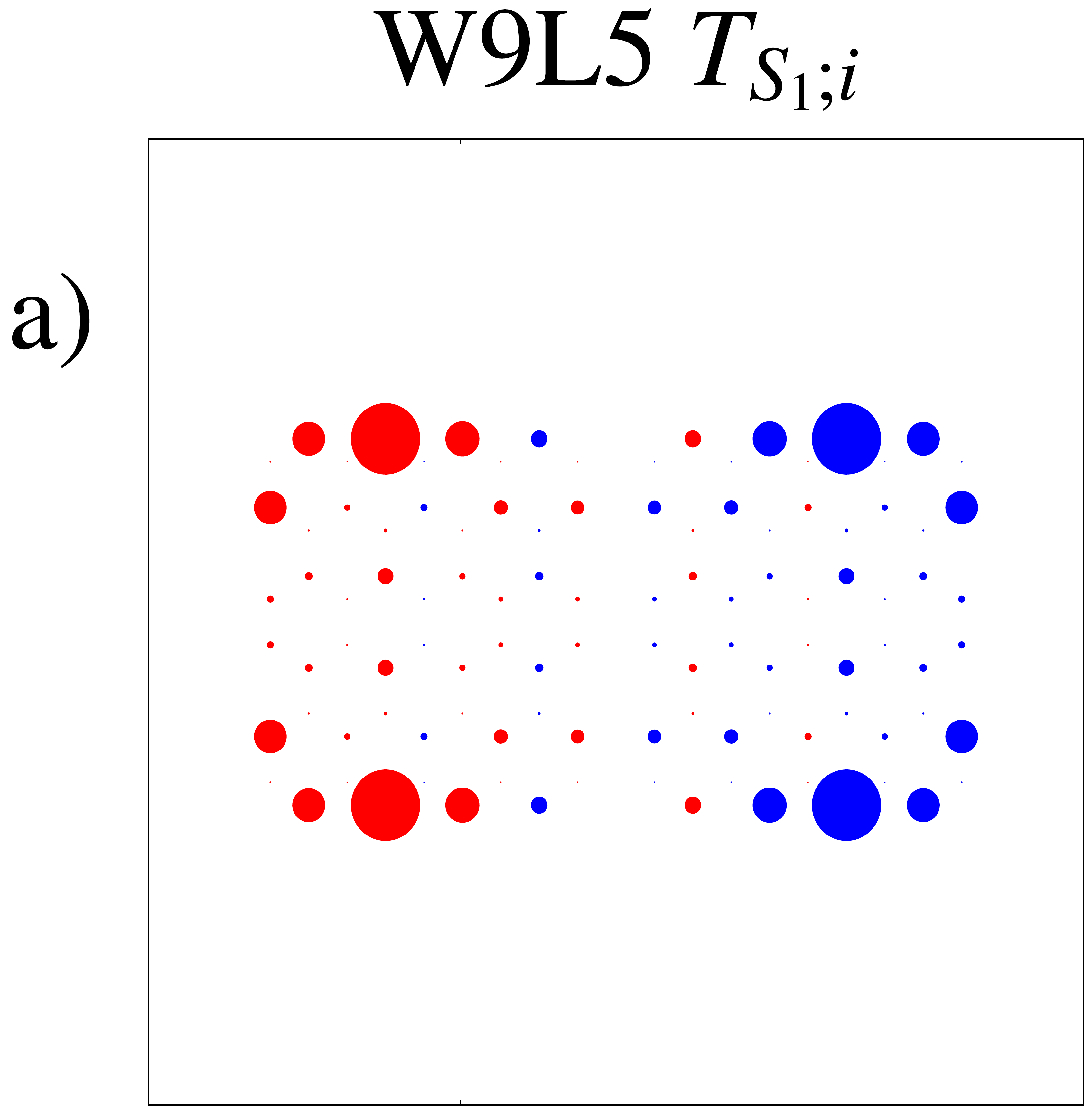} &
\includegraphics[width=0.4\columnwidth]{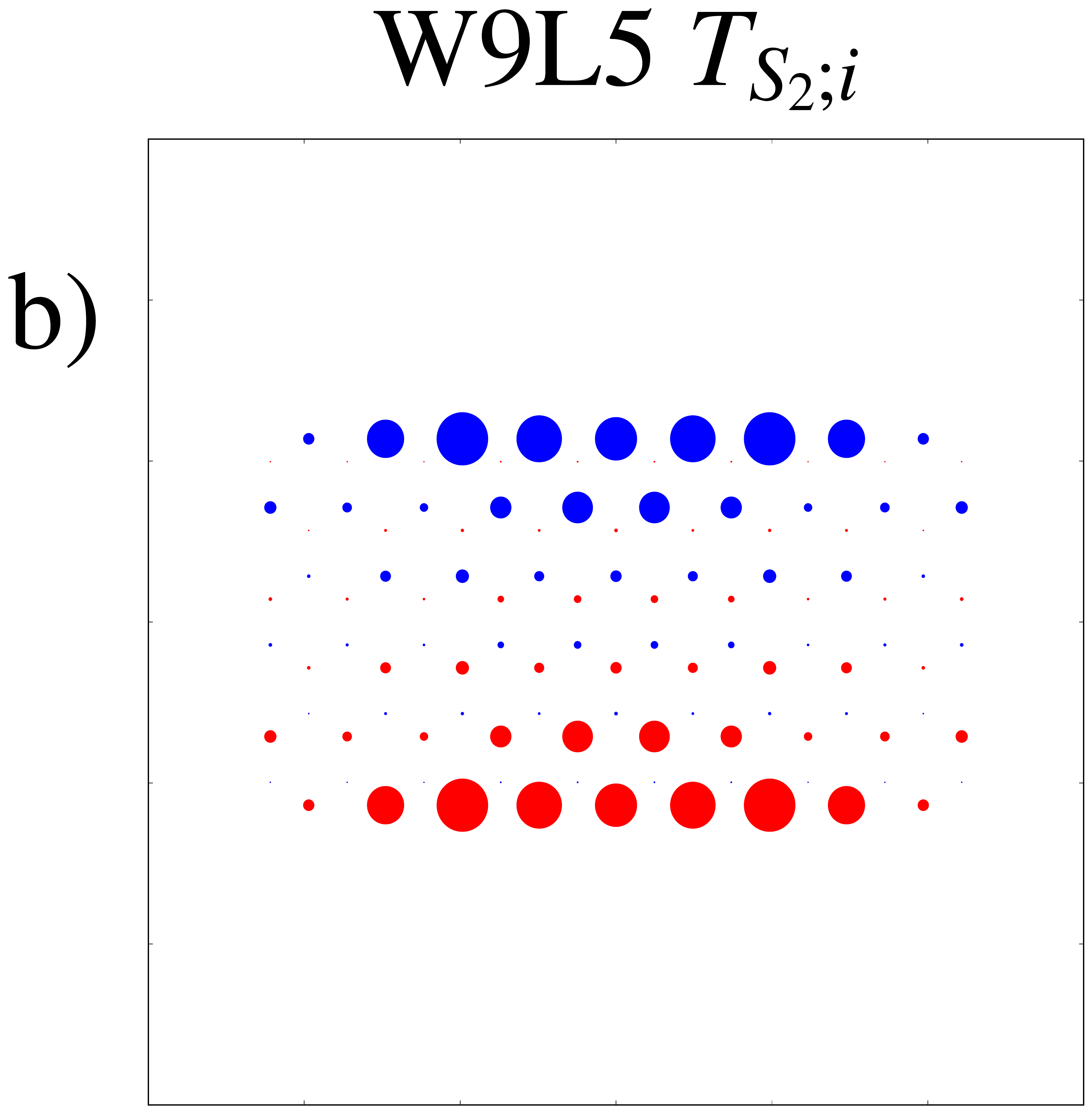} 
\end{tabular}
\end{tabular}
\caption{(Top) Absorption spectrum of the W$n$L5 family of flakes. As we consider larger flakes, the first two absorption peaks are red shifted. For the largest flake, W9L5, the lowest energy peak corresponds to a transition with a transition dipole moment oriented along the long axis (here $\hat{\mbf{x}}$) of the system, while the second lowest energy peak is attributed to a relatively weak transition with a transition dipole moment oriented along the short (here $\hat{\mbf{y}}$) axis of the flake. (Bottom) Plot of a) $T_{S_{1};i}$ and b) $T_{S_{2};i}$ for the W9L5 system. The electrons are concentrated on the zig-zag edges for both these transitions.}
\label{fig4}
\end{figure}

In Fig. \ref{fig2}, we plot the absorption spectrum for the W$n$L3 family of flakes. Again, as the flake gets larger, the first absorption peak is red shifted.  For larger flakes, the first absorption peak is associated with a transition with a transition dipole moment  polarized along the long axis (here $\hat{\mbf{x}}$) of the flake. 
We also plot $T_{S_{1};i}$ for the W11L3 flake in Fig. \ref{fig2}. 
For the first bright excited state, $T_{S_{1};i}$ indicates that the electrons involved in that particular transition are localized on the zig-zag edges. 
The first bright excited state is composed of several HF double excitations.  These states become bright because the CI ground state of these flakes has significant contributions from HF double excitations, which leads to a non-zero transition dipole moment between the ground state and the $S_{1}$ state. 
Unlike the W3L$n$ family of flakes, the state which predominantly involves the HF single excitation between edge modes, the $S_{2}$ state in the W$n$L3 family, is weakly bright and higher in energy than the first excited state. 
This optical transition is weak because the dominant contribution to the excited state, the HF single excitation which involves the transition from the edge modes, has a transition dipole moment which is oriented on the short axis (here $\hat{\mbf{y}}$) of the flake. The quantity $T_{S_{2};i}$ is plotted in Fig. \ref{fig2}; it shows significant electron concentration on the zig-zag edges of the system. In the W3L$n$ family,  the $S_{1}$ state is composed primarily of HF single excitations, while the $S_{2}$ state is composed mainly of HF double excitations. However, in the W$n$L3 family this trend is reversed. These features cannot be observed using mean-field calculations only \cite{J.A.Verges2015}, and require the CI calculation. 

We now turn to the W5L$n$ family of flakes. 
In Fig. \ref{fig3}, we plot the absorption spectrum for the W5L$n$ family of flakes. 
Both the first absorption peak, associated with a very weak transition whose transition dipole moment is polarized along the long axis (here $\hat{\mbf{y}}$) of the flake, and the second and much stronger absorption peak, associated with a transition whose transition dipole moment is polarized along the short axis (here $\hat{\mbf{x}}$), are red shifted as the flake gets larger. 
We plot $T_{S_{1};i}$  in Fig. \ref{fig3}. For this particular system,  $T_{S_{1};i}$ extends around the entire flake, although with a significant contribution from the zig-zag edges. 
We also plot $T_{S_{2};i}$ in Fig. \ref{fig3}, which shows that the electrons are concentrated almost exclusively on the zig-zag edges of the system for this transition. 
In these flakes, the $S_{1}$ state is primarily composed of HF single excitations involving transitions from the edge modes, while for larger flakes, the $S_{1}$ state also has significant contributions from HF single excitations involving the bulk modes.
 This mixing is larger in these flakes than in the W3L$n$ family. The $S_{2}$ state is primarily composed of HF double excitations. 
 %The $S_{2}$ state is primarily composed of HF double excitations (\ref{hfdoubs}). % ??
 The W5L$n$ family of flakes behaves similarly to the W3L$n$ family of flakes, except in the strength of the first absorption peak. %This weak absorption peak is associated with a transition from the ground state to a state composed primarily of HF single excitations of edge modes.
 This weak absorption peak, associated with a transition whose transition dipole moment is polarized along the long axis of the flake, is dwarfed in intensity by the bright peak  associated with a transition whose transition dipole moment is polarized along the short axis.

\begin{figure}[tb!]
\begin{center}
\includegraphics[scale=0.2]{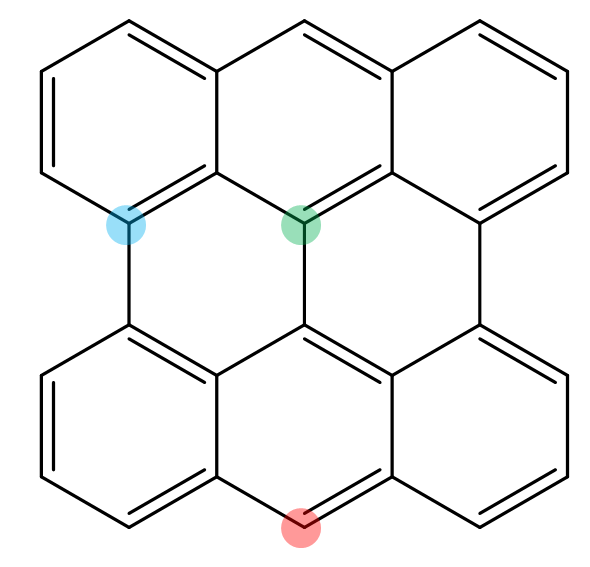}
\caption{Cartoon of a W3L3 flake showing the locations of various impurity potentials. The red circle represents an impurity placed on the zig-zag edges, the light blue circle represents an impurity placed on the arm-chair edges, and the green circle represents an impurity placed in the middle, or so-called ``bulk'' region, of the flake.}
\label{figimpcartoon}
\end{center}
\end{figure}

\begin{figure}[htb!]
\begin{center}
\includegraphics[width=1.0\columnwidth]{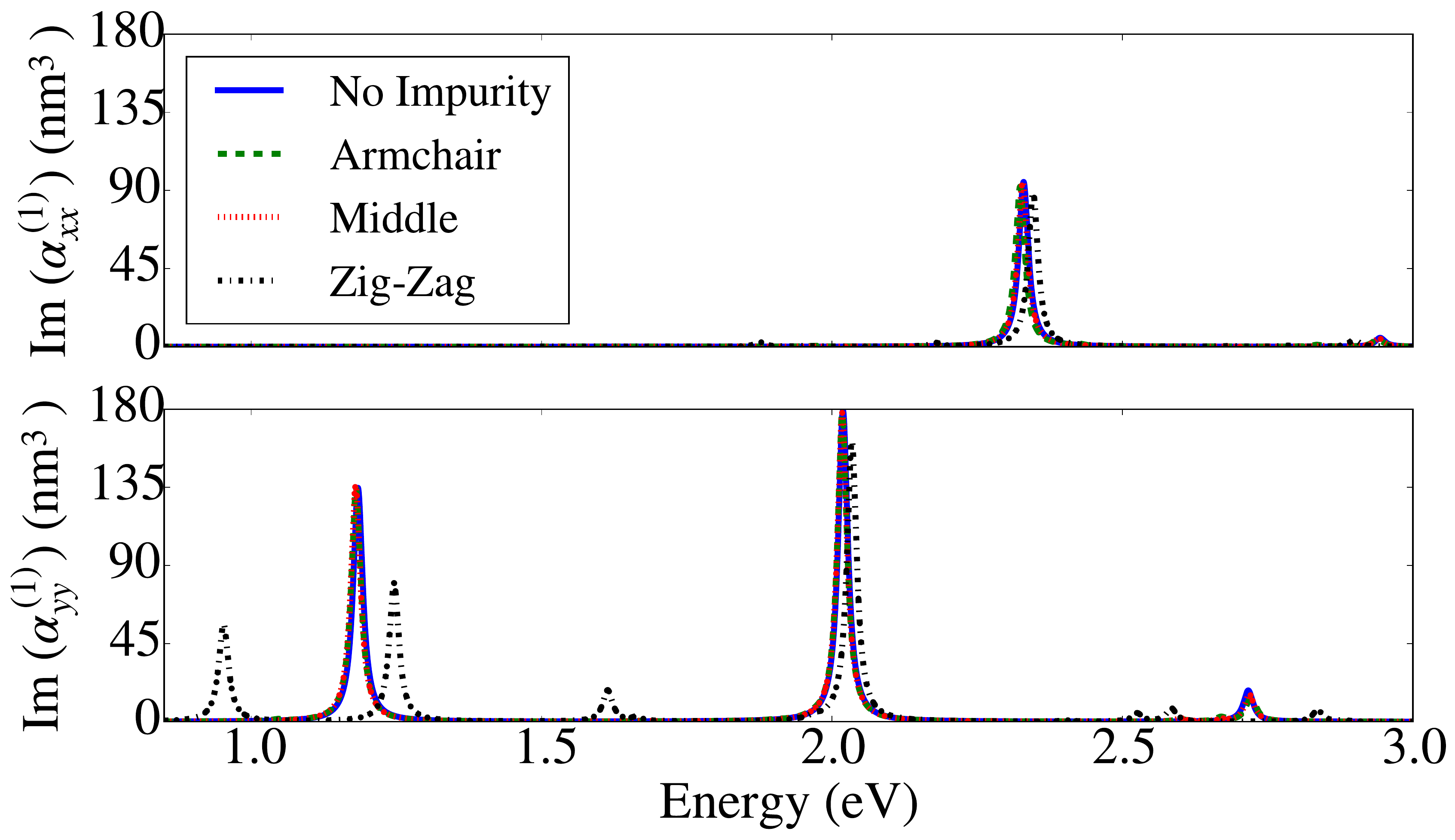}
\caption{Absorption spectrum of the W3L11 flake with varied impurity locations. The absorption remains unchanged for impurity potentials centered on the armchair edges or in the middle of the flake, but placing an impurity potential on one of the zig-zag edges results in two new low energy peaks, one red shifted, one blue shifted from the original absorption line. The parameters used to model the impurity potential were $\overline{\varepsilon}_{max} = t/3$ and $\tau = l_{b}$.}
\label{fig5}
\end{center}
\end{figure}

\begin{figure}[htb!]
\begin{tabular}{c}
\includegraphics[width=0.4\columnwidth]{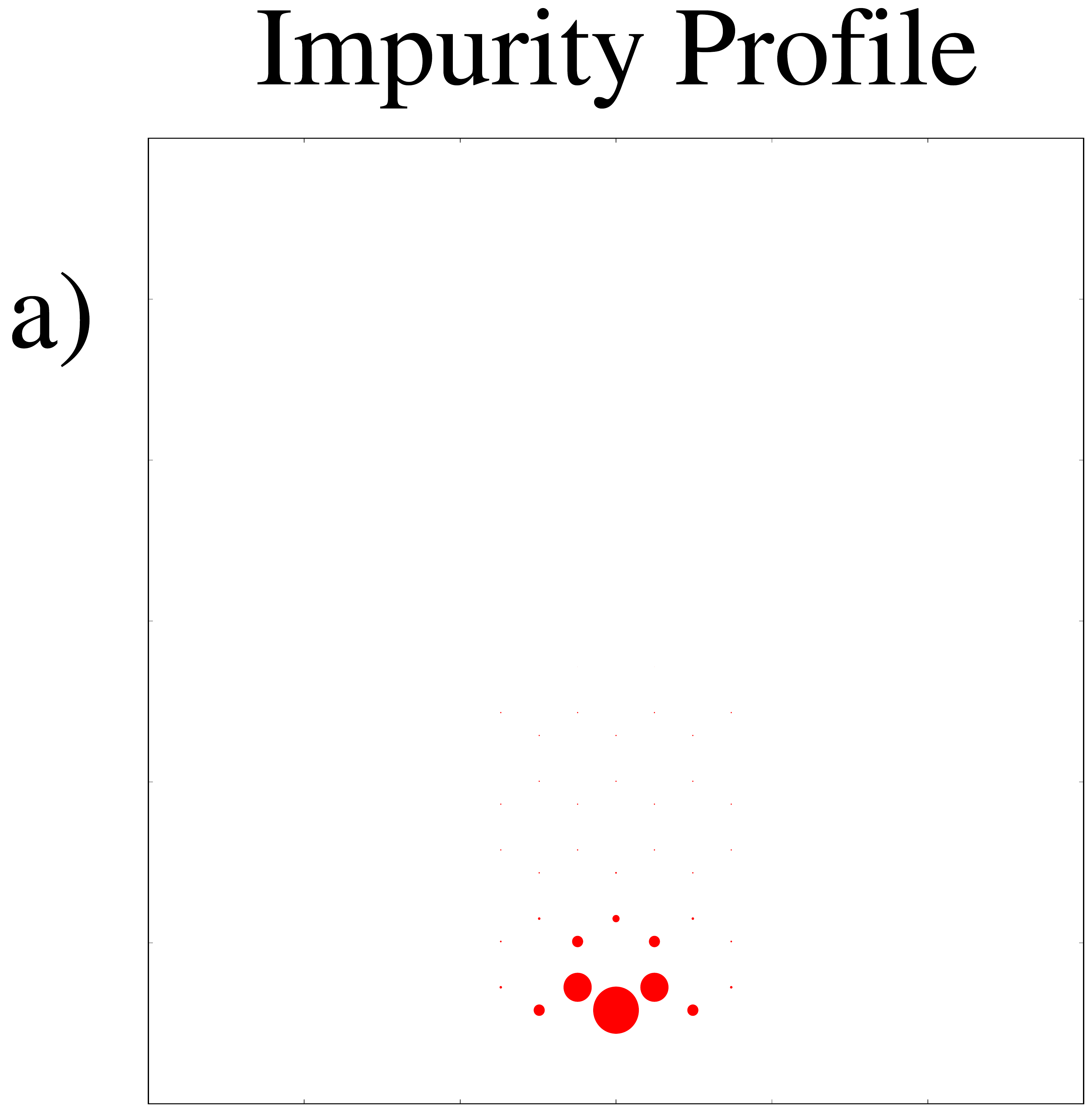} \\
\begin{tabular}{rr}
\includegraphics[width=0.45\columnwidth]{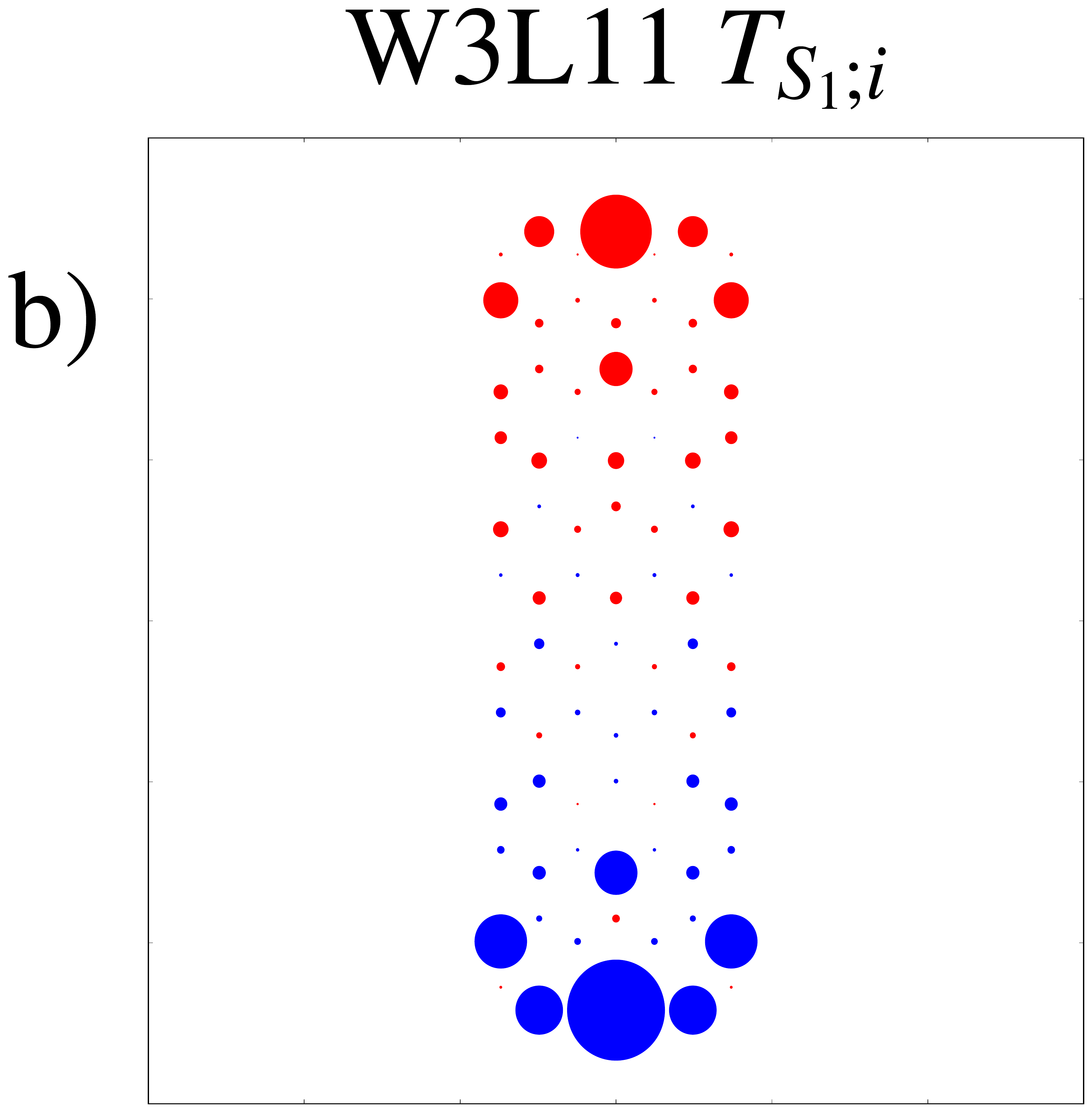} &
\includegraphics[width=0.45\columnwidth]{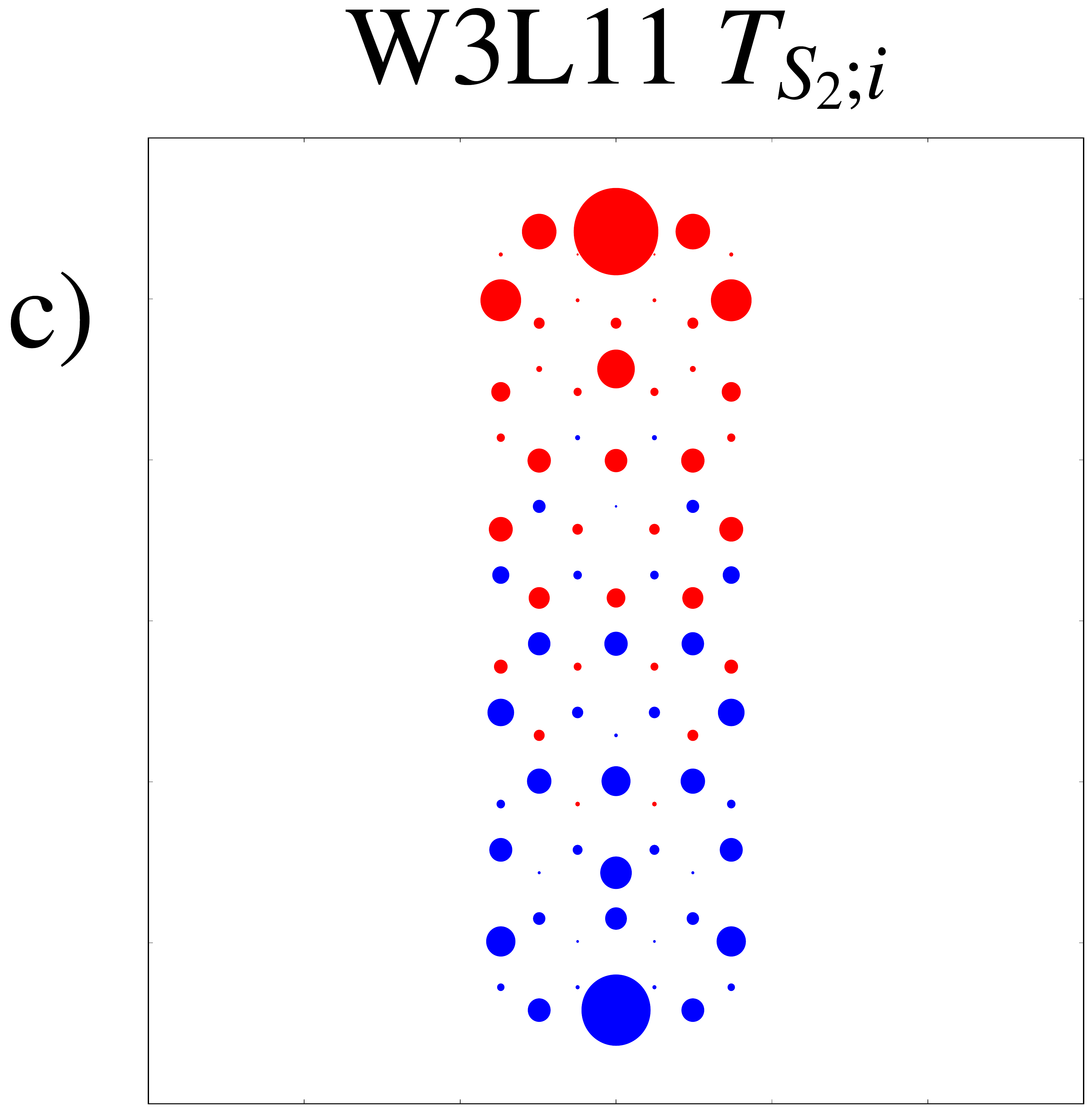}
\end{tabular}
\end{tabular}
\caption{a) Profile of an impurity placed in the zig-zag bottom site of the W3L11 flake. Plot of b) $T_{S_{1};i}$ and c) $T_{S_{2};i}$ for the W3L11 system with an impurity potential from a). The charge distributions for these transitions are still concentrated around the zig-zag edges. The parameters used to model the impurity potential were $\overline{\varepsilon}_{max} = t/3$ and $\tau = l_{b}$. }
\label{fig6}
\end{figure}

\begin{figure}[htb!]
\begin{tabular}{c}
\includegraphics[width = 1.0\columnwidth]{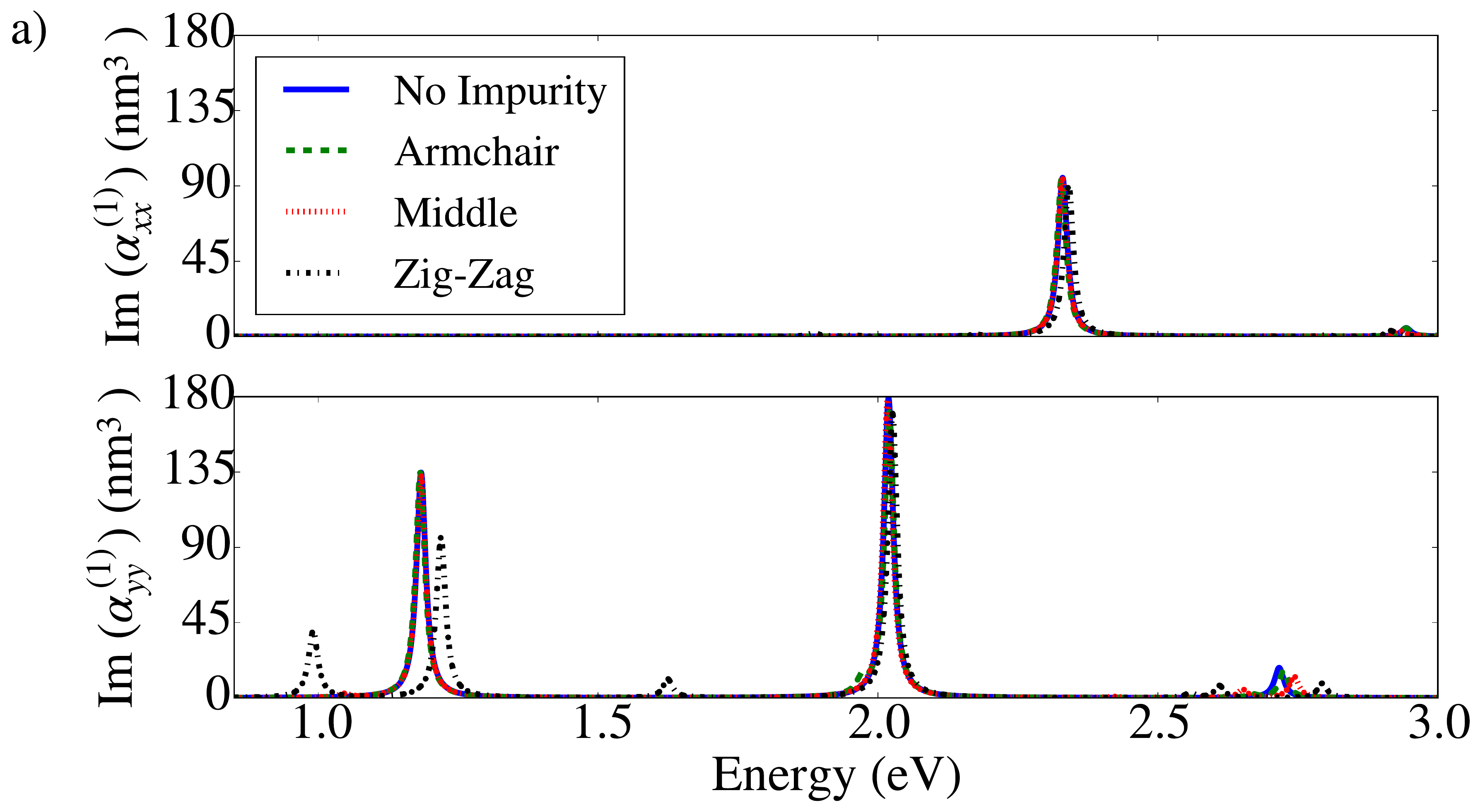} \\
\includegraphics[width = 1.0\columnwidth]{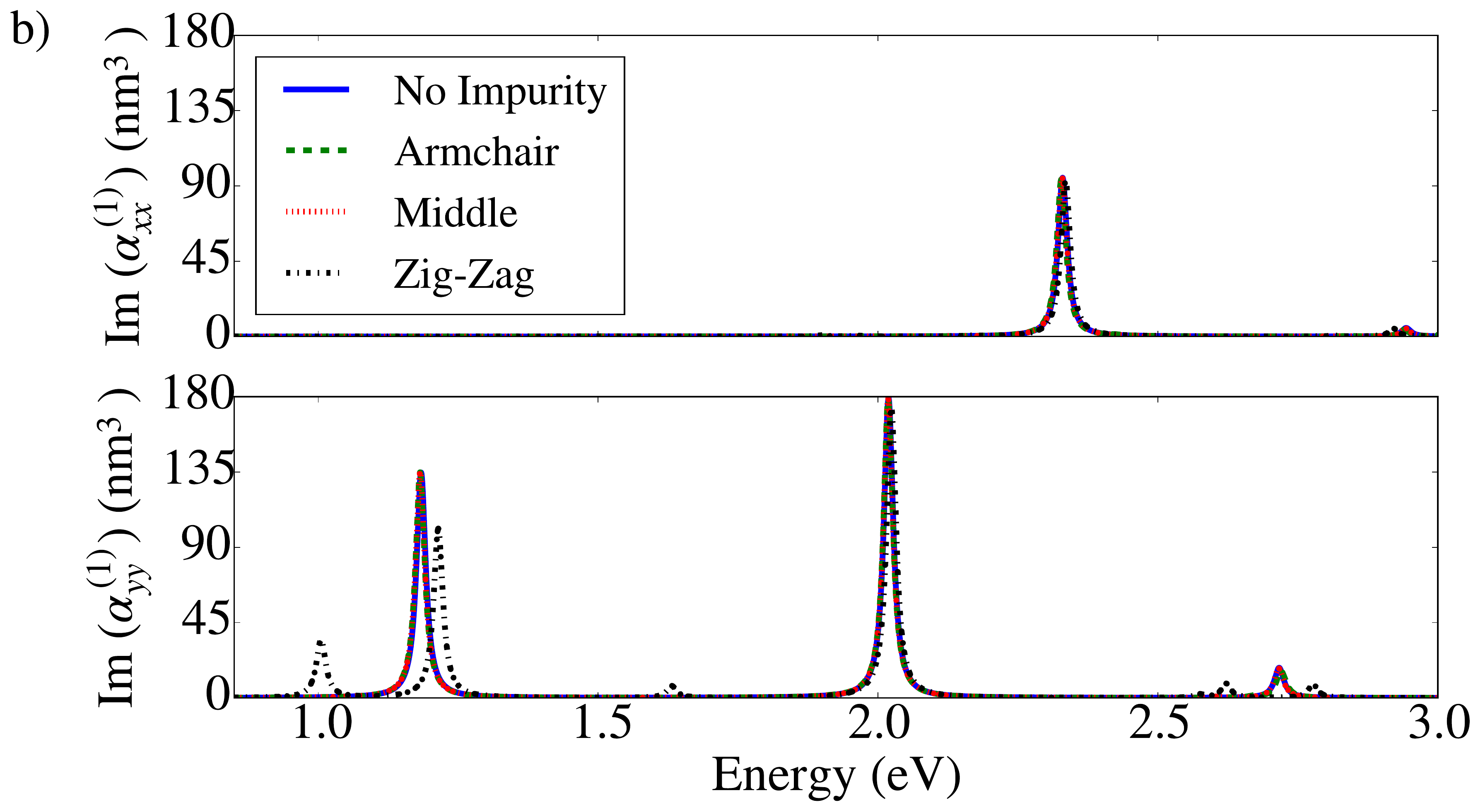}
\end{tabular}
\caption{Absorption spectrum of the W3L11 flake with varied impurity locations, impurity strengths, and ranges. Plot of a) absorption spectrum for a shorter range impurity potential localized to a particular site with $\overline{\varepsilon}_{max} = t/3$ and $\tau = l_{b}/5.0$, and b) the absorption spectrum with a weaker impurity potential, with parameters of $\overline{\varepsilon}_{max} = t/5$ and $\tau = l_{b}$.}
\label{fig5a}
\end{figure}

Finally we investigate the W$n$L5 family of flakes. We plot the absorption spectrum for these flakes in Fig \ref{fig4}. As the flake gets larger, the first absorption peak is again red shifted, but unlike the W5L$n$ family of flakes, the lowest absorption peak is very strong. The lowest absorption peak corresponds to the ground to $S_{1}$ transition, the associated transition dipole moment is polarized along the long axis (here $\hat{\mbf{x}}$) of the flake. The second lowest energy absorption peak corresponds to the ground to $S_{2}$ transition, the associated transition dipole moment is polarized along the short axis (here $\hat{\mbf{y}}$) of the flake. We plot $T_{S_{1};i}$ and $T_{S_{2};i}$ for the W9L5 flake in Fig \ref{fig4}. 
For both these states, most of the electron concentration is located on the zig-zag edges. Unlike the W5L$n$ family of flakes, the $S_{1}$ state in the W$n$L5 family is composed mainly of several HF double excitations, while the $S_{2}$ state is composed primarily of HF single excitations involving transitions between edge modes, but with significant mixing from HF single excitations involving transitions between bulk modes. 
The trends we see in the W$n$L5 family are similar to the W$n$L3 family, for which the lowest energy bright state is composed of mainly HF double excitations while the second lowest energy bright state is composed mainly of HF single excitations. 

In summary, for the pristine GFs the lowest energy bright transition invariably has a transition dipole moment which is polarized along the long axis of the flake. When the armchair edges are larger than the zig-zag edges, the lowest energy bright excited state is composed mainly of HF single excitations. However, when the zig-zag edges are larger than the armchair edges, the lowest energy bright excited state is composed mainly of HF double excitations. 
\section{\label{sec-imp} The effects of impurities on the  absorption spectra}

\begin{figure}[htb!]
\begin{center}
\includegraphics[width=1.0\columnwidth]{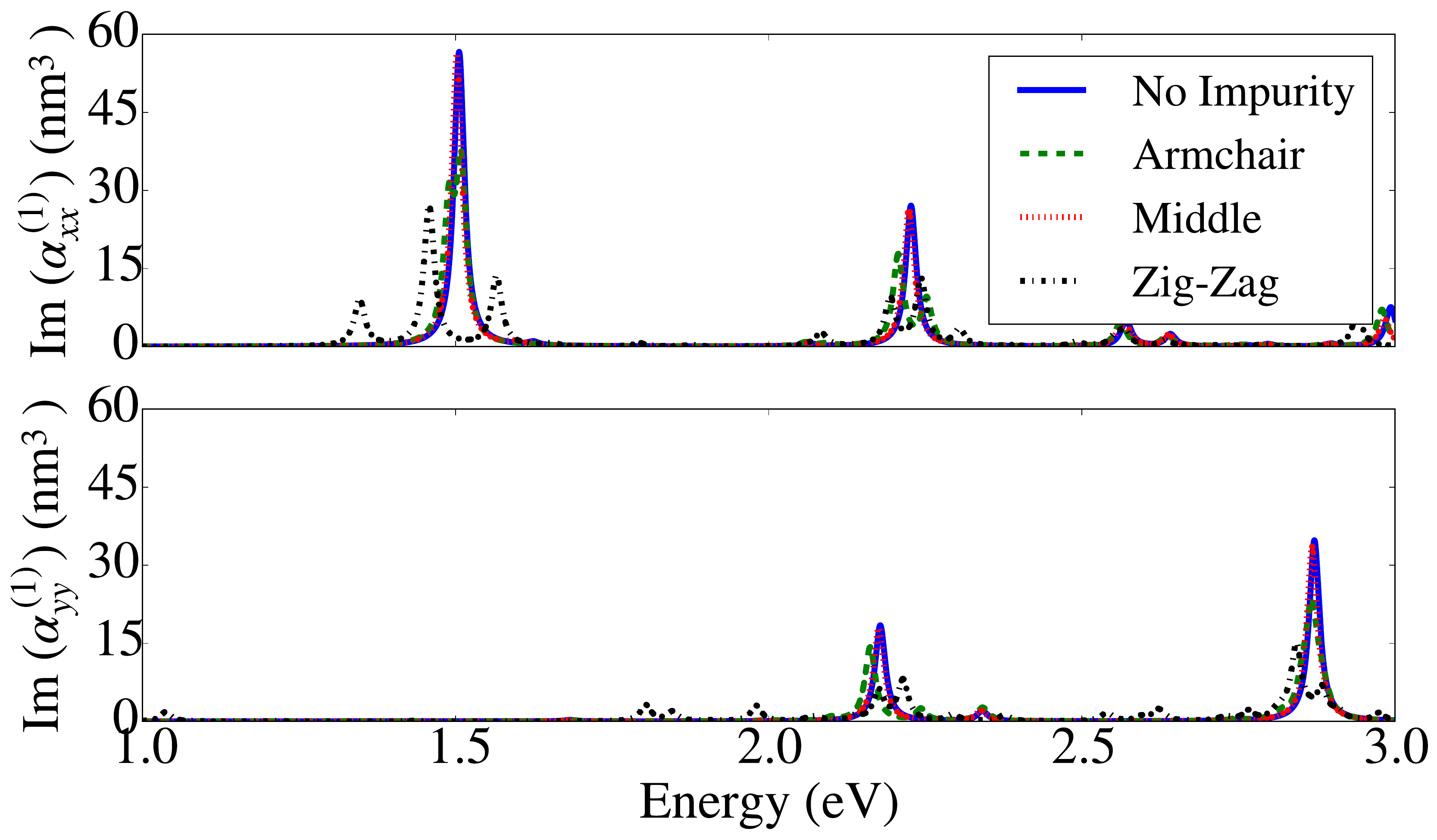}
\caption{Absorption spectrum of the W11L3 flake with varied impurity locations. The absorption remains relatively unchanged for impurities located in the middle of the flake, but placing an impurity potential on one of the zig-zag edges results in several new peaks. The zig-zag impurity potential also induces a very weak, low energy absorption with a transition dipole moment oriented in the $\hat{\mbf{y}}$ direction, the short axis of the flake. The parameters used to model the impurity potential were $\overline{\varepsilon}_{max} = t/3$ and $\tau = l_{b}$.}
\label{fig7}
\end{center}
\end{figure}

\begin{figure*}[htb!]
\begin{tabular}{rrr}
\includegraphics[width = 0.19\textwidth]{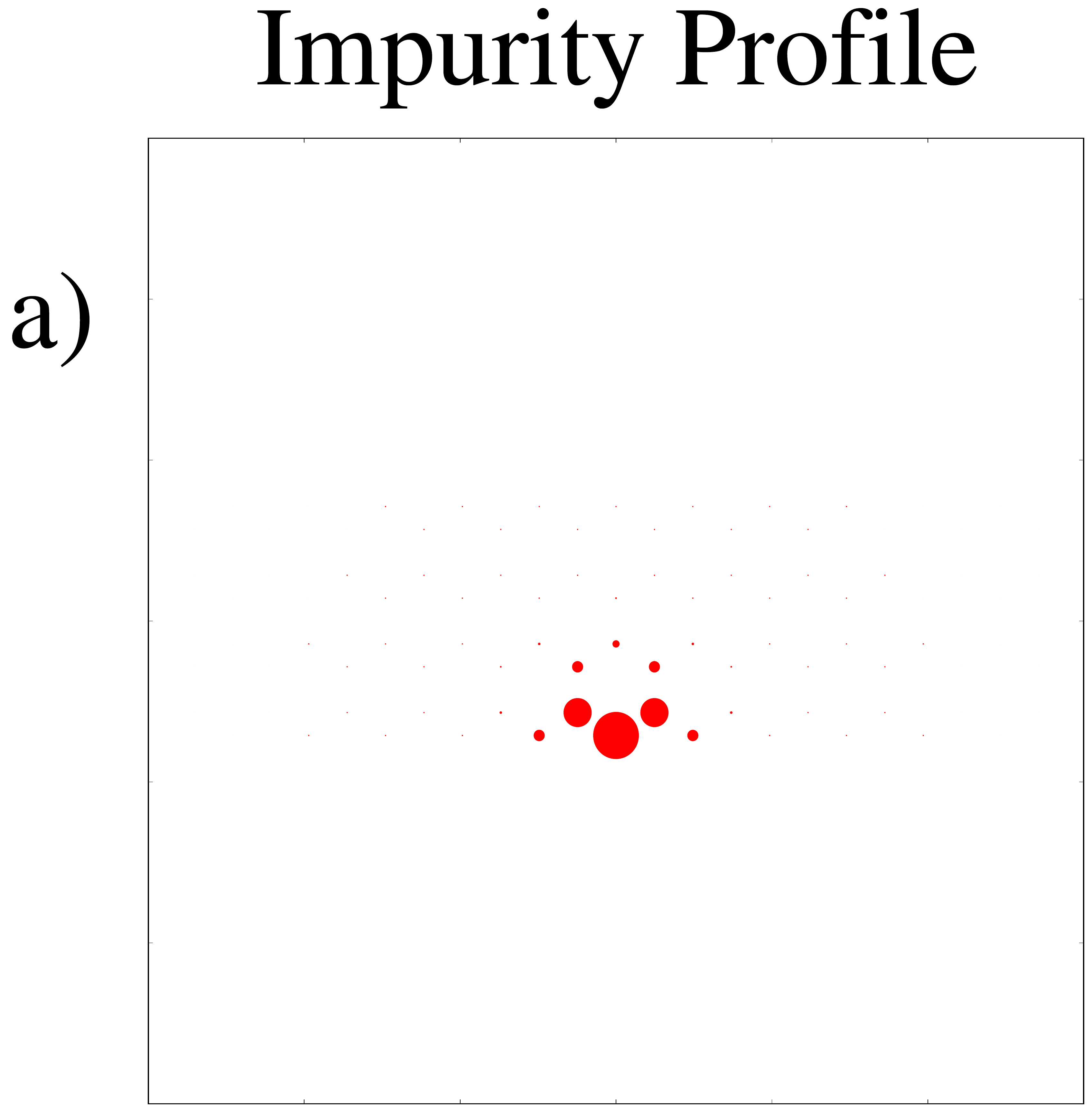} &
\includegraphics[width=0.19\textwidth]{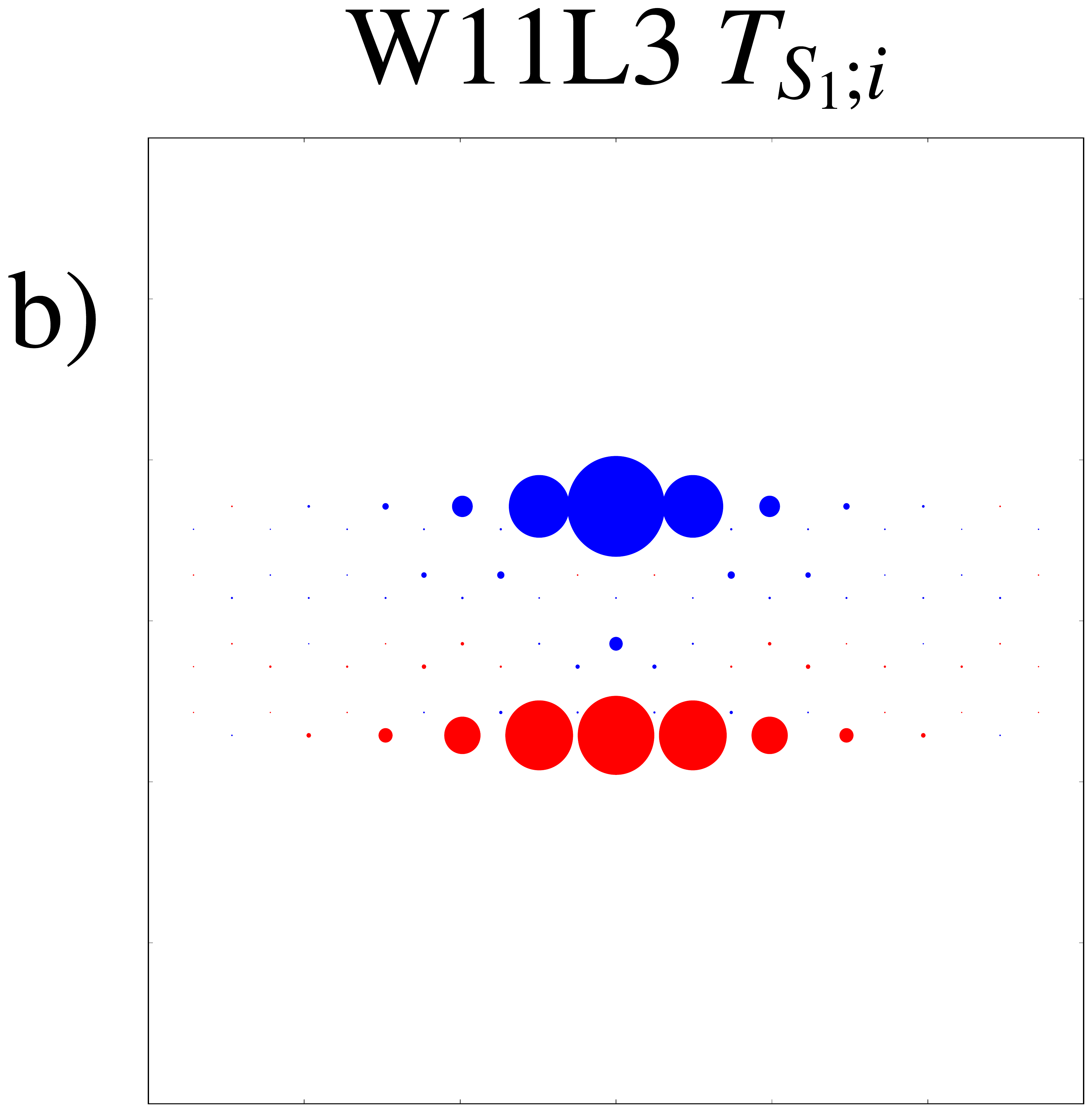} &
\includegraphics[width=0.19\textwidth]{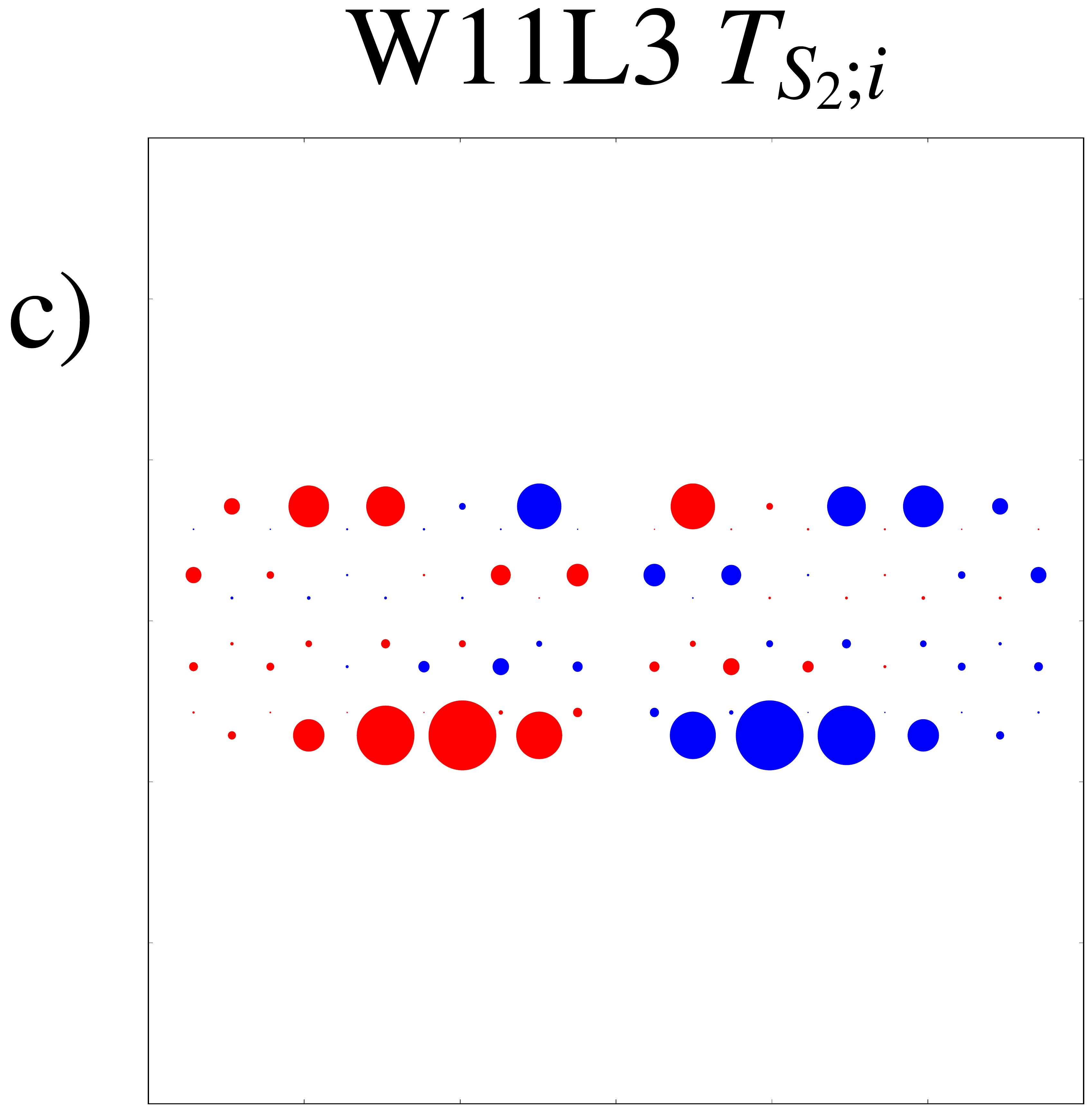} 
\end{tabular}
\begin{tabular}{rr}
\includegraphics[width=0.19\textwidth]{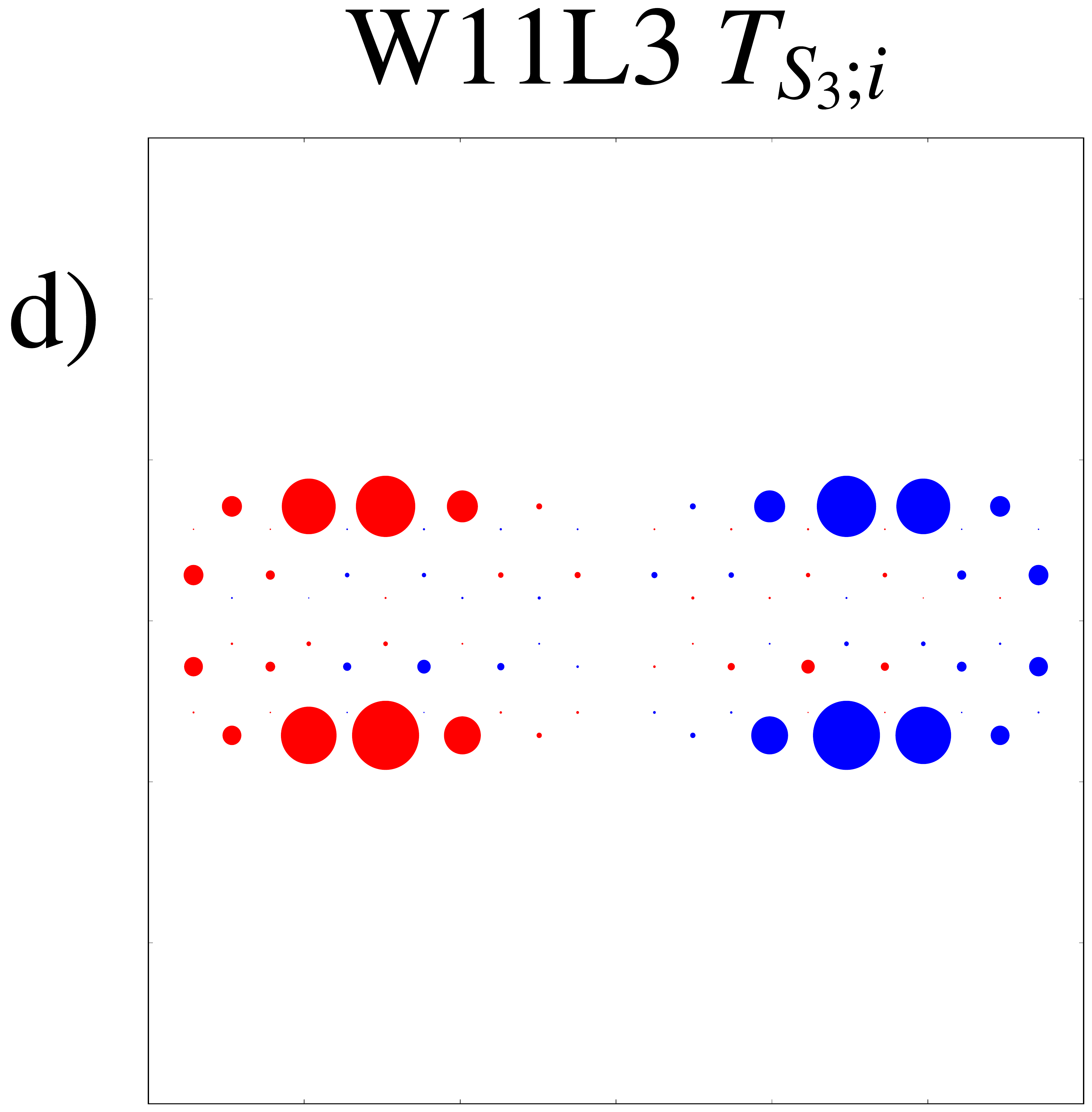} &
\includegraphics[width=0.19\textwidth]{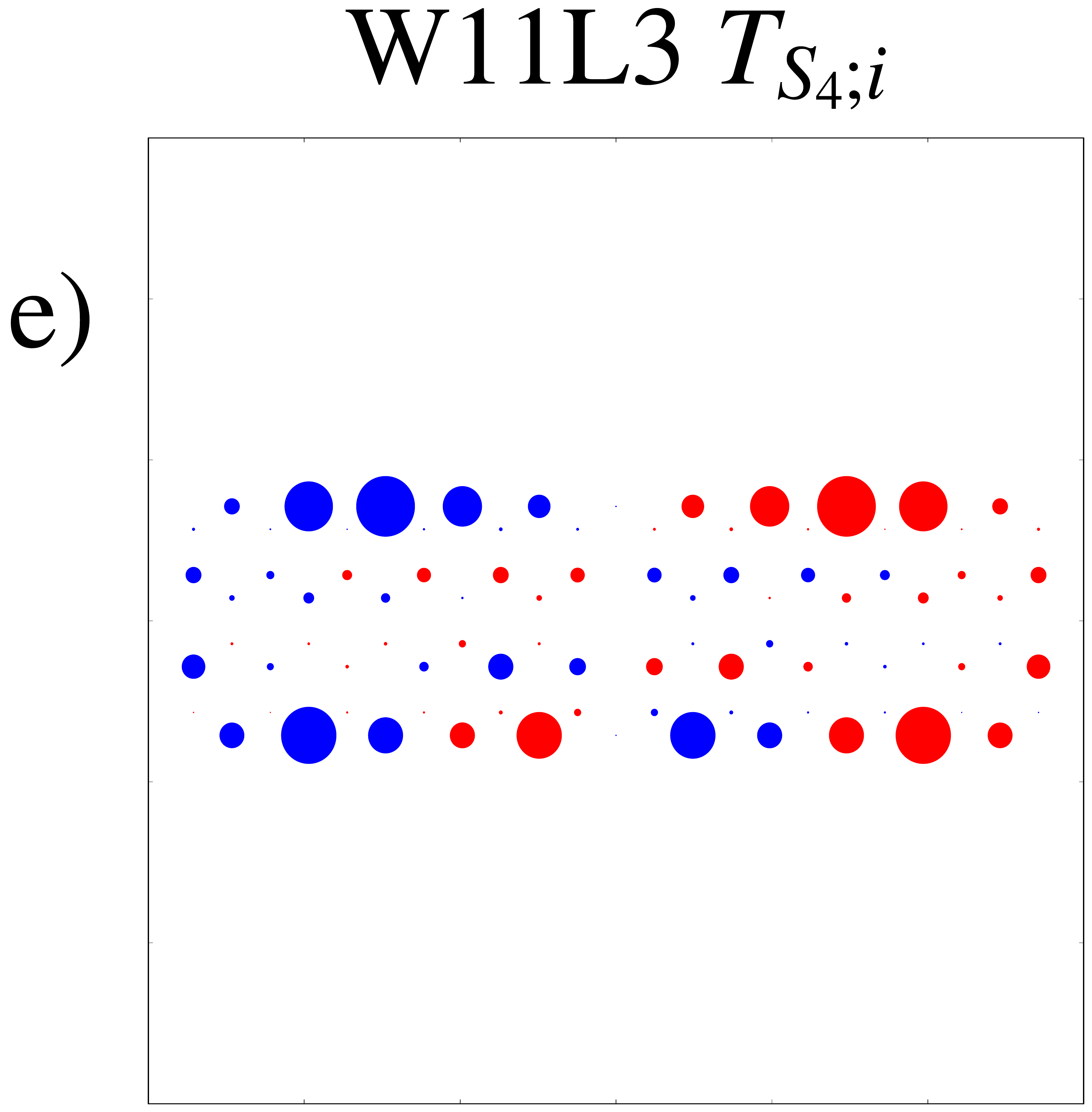}
\end{tabular}
\caption{Plot of the a) profile of an impurity placed in the zig-zag bottom site of the W11L3 flake with the same parameters as in Fig \ref{fig7}. Plot of b) $T_{S_{1};i}$, c) $T_{S_{2};i}$, d) $T_{S_{3};i}$, and e) $T_{S_{4};i}$ for the W11L3 flake with an impurity potential shown in a). The charge distributions are still concentrated on the zig-zag edges for these transitions.}
\label{fig8}
\end{figure*}

\begin{figure}[htb!]
\begin{tabular}{c}
\includegraphics[width = 1.0\columnwidth]{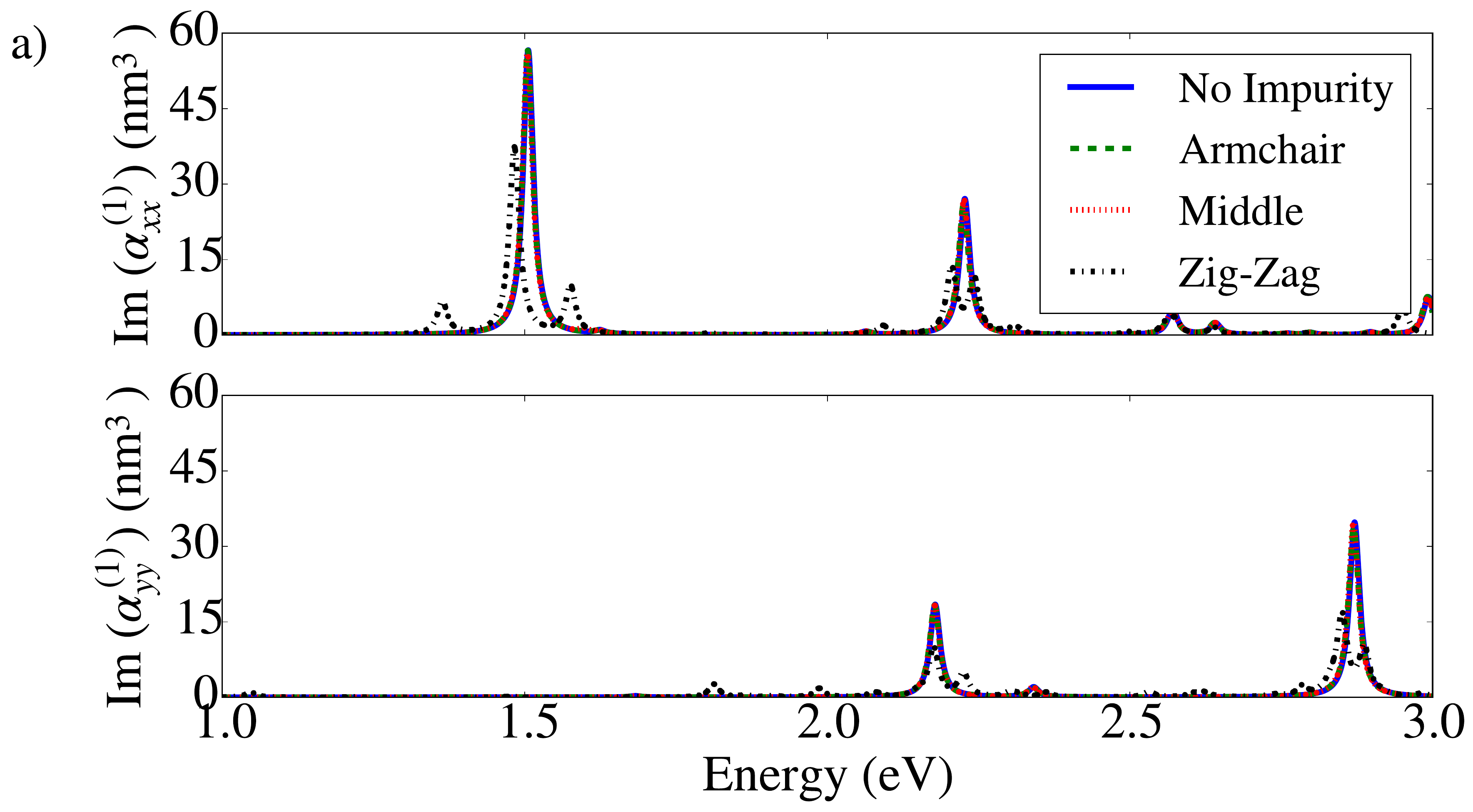} \\
\includegraphics[width = 1.0\columnwidth]{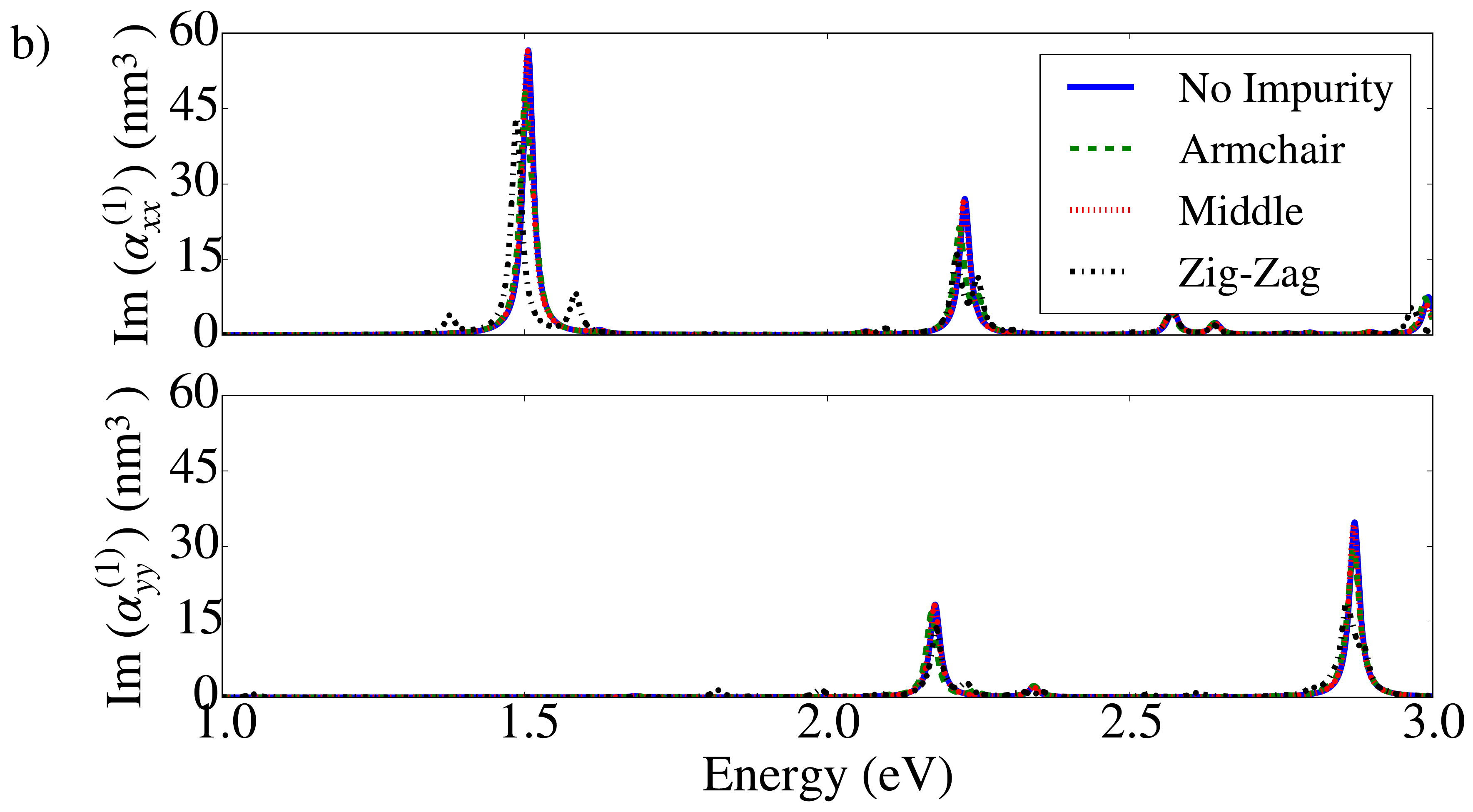}
\end{tabular}
\caption{Absorption spectrum of the W11L3 flake with varied impurity locations, varied impurity strengths, and ranges. Plot of a) the absorption spectrum of a W11L3 flake with a shorter range impurity potential, parameters used to model the impurity were $\overline{\varepsilon}_{max} = t/3$ and $\tau = l_{b}/5$ and b) the absorption spectrum of a W11L3 flake with a weaker impurity potential, parameters used to model the weak impurity potential were $\overline{\varepsilon}_{max} = t/5$ and $\tau = l_{b}$.}
\label{fig7a}
\end{figure}

Impurities can impact the optical properties of GFs depending on the location, strength, and range of influence of the impurity potentials \cite{C.H.A.Wong2014, M.Buzaglo2016, L.Liao2010,Y.M.Lin2010}.  
In this section, we analyze the effect of impurities on the optical properties of the GF families discussed in the previous section. 
We consider impurity potentials centered on the zig-zag edges, armchair edges, and in the center of the flake, and compute the optical properties of the GF. 
We illustrate these impurity potentials in Fig. \ref{figimpcartoon}. 
We show results for positive impurity potentials. Changing the sign of the impurity potential does not qualitatively change the results presented in this section. 
This can be understood by treating the impurity potential perturbatively. The first order correction to the energy of states, given by $\la \psi | H_{imp} | \psi \ra$  where $\psi$ corresponds to a particular CI state, is similar for the ground state and states close in energy to it.
This is due to the fact that the densities associated with the ground state and the low energy unperturbed states are approximately similar around the impurity location. 
Therefore, the effect of impurity potentials becomes significant only at second order in the perturbation, for which the energy corrections are independent of the sign of the impurity potential.

We first examine a flake with larger armchair edges than zig-zag, namely the W3L11 flake. 
The absorption spectrum of the W3L11 flake, with a Gaussian impurity potential located on the middle, armchair, and zig-zag edges of the flakes are plotted in Fig. \ref{fig5}.
We use the impurity potential parameters $\overline{\varepsilon}_{max} = t/3$ and $\tau = l_{b}$, for which the absorption spectra of the flakes with the impurity on its armchair edge or in the center of the flake are unchanged from that of the pristine GF. %\hl{We point out that due to the off center location of the armchair impurity, it introduces a very weak cross polarization appearing in the} $\text{Im } \left( \alpha^{(1)}_{xy} \right)$ and $\text{Im } \left( \alpha^{(1)}_{yx} \right)$.
The low-lying bright excited states are not significantly affected by the presence of the impurity potential on these locations, as evidenced by the absorption spectrum and the joint density of states. This is because the spatial profiles of the low lying excitations have little electron concentration on the armchair edges or in the middle of the flake. %We also point out that pristine rectangular GFs have transition dipole moments polarized along their long axis, but a zig-zag impurity induces very weak cross polarization appearing in the $\text{Im} (\alpha^{(1)}_{ij})$, for $i \neq j$.
As illustrated in the previous section, the low energy transitions in rectangular GFs have electron concentration on the zig-zag edges.
The electron concentration on the zig-zag edges is unaffected by impurity potentials centered on the middle or armchair edges of the flake, unless the impurity potentials extend to the zig-zag edges.

In contrast, an impurity on a zig-zag edge of a GF can have a significant impact on its optical properties. This is because the charge distributions involved in the bright transitions have significant concentration on the zig-zag edges.
The impurity blue shifts the first absorption peak, as the excited state involved in the transition becomes less energetically favorable due to the presence of the impurity potential. It also mixes a dark transition with a bright HF single excitation that involves an excitation of an electron between the edge modes. 
In Fig. \ref{fig6}, we show the quantities $T_{S_{1};i}$ and $T_{S_{2};i}$ for an impurity potential centered on a zig-zag edge of the flake.
%The presence of the impurity potential changes the sign of the dipole moment. 
%We also point out that pristine rectangular GFs have transition dipole moments polarized along their long axis, but a zig-zag impurity induces very weak cross polarization appearing in the $\text{Im} (\alpha^{(1)}_{ij})$, for $i \neq j$.

\begin{figure}[htb!]
\begin{center}
\includegraphics[width=1.0\columnwidth]{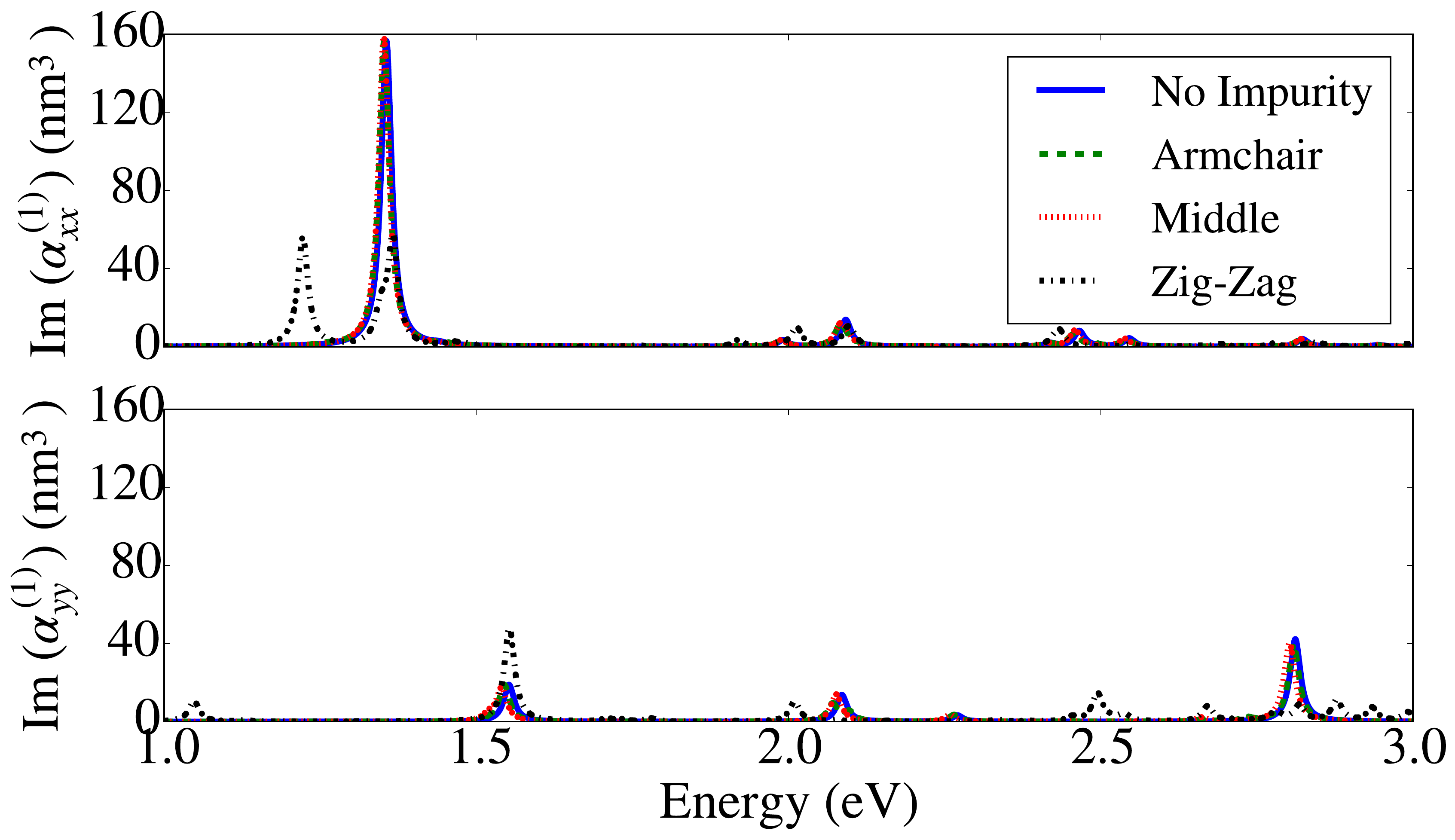}
\caption{Absorption spectrum of the W9L5 flake with varied impurity locations. The absorption remains relatively unchanged for impurities located on either the armchair edges or in the middle of the flake. However, placing an impurity potential on one of the zig-zag edges results in several new peaks, including a relatively weak low energy absorption which has a transition dipole moment that is polarized in the $\hat{\mbf{y}}$ direction, the short axis of the flake. The parameters used to model the impurity potential were $\overline{\varepsilon}_{max} = t/3$ and $\tau = l_{b}$.}
\label{fig9}
\end{center}
\end{figure}
\begin{figure*}[htb!]
\begin{tabular}{rrr}
\includegraphics[width=0.19\textwidth]{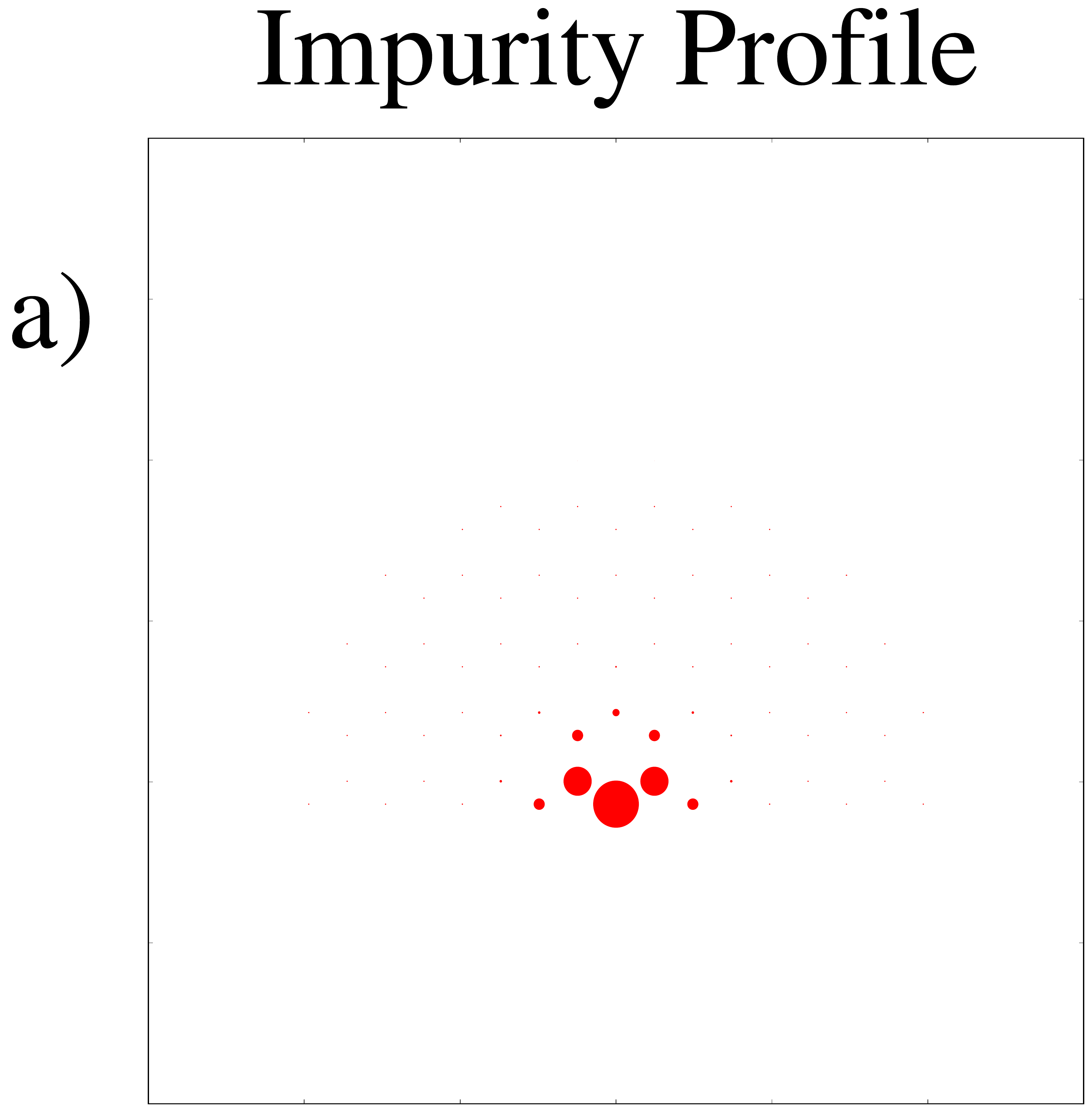}
\includegraphics[width=0.19\textwidth]{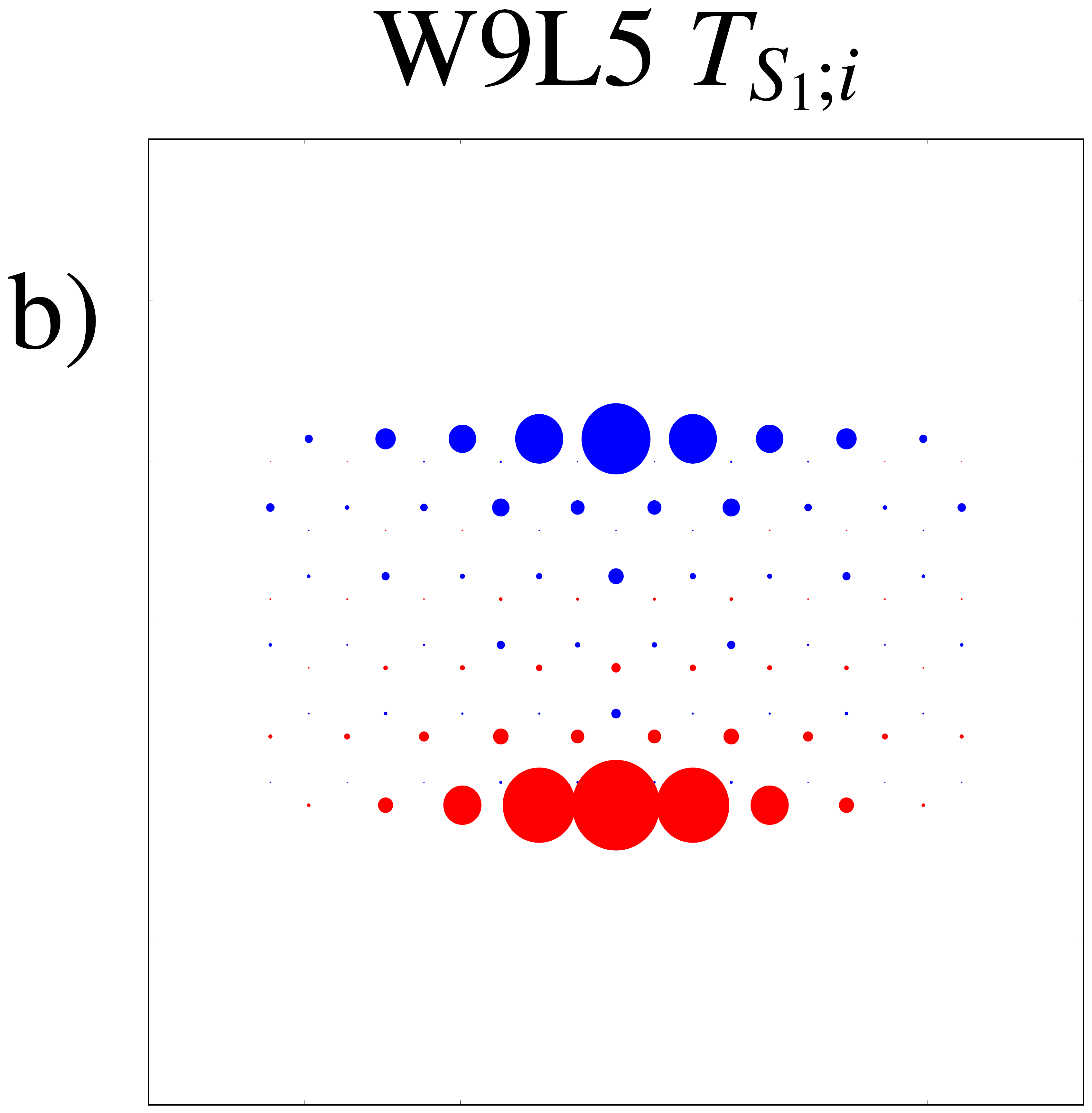} &
\includegraphics[width=0.19\textwidth]{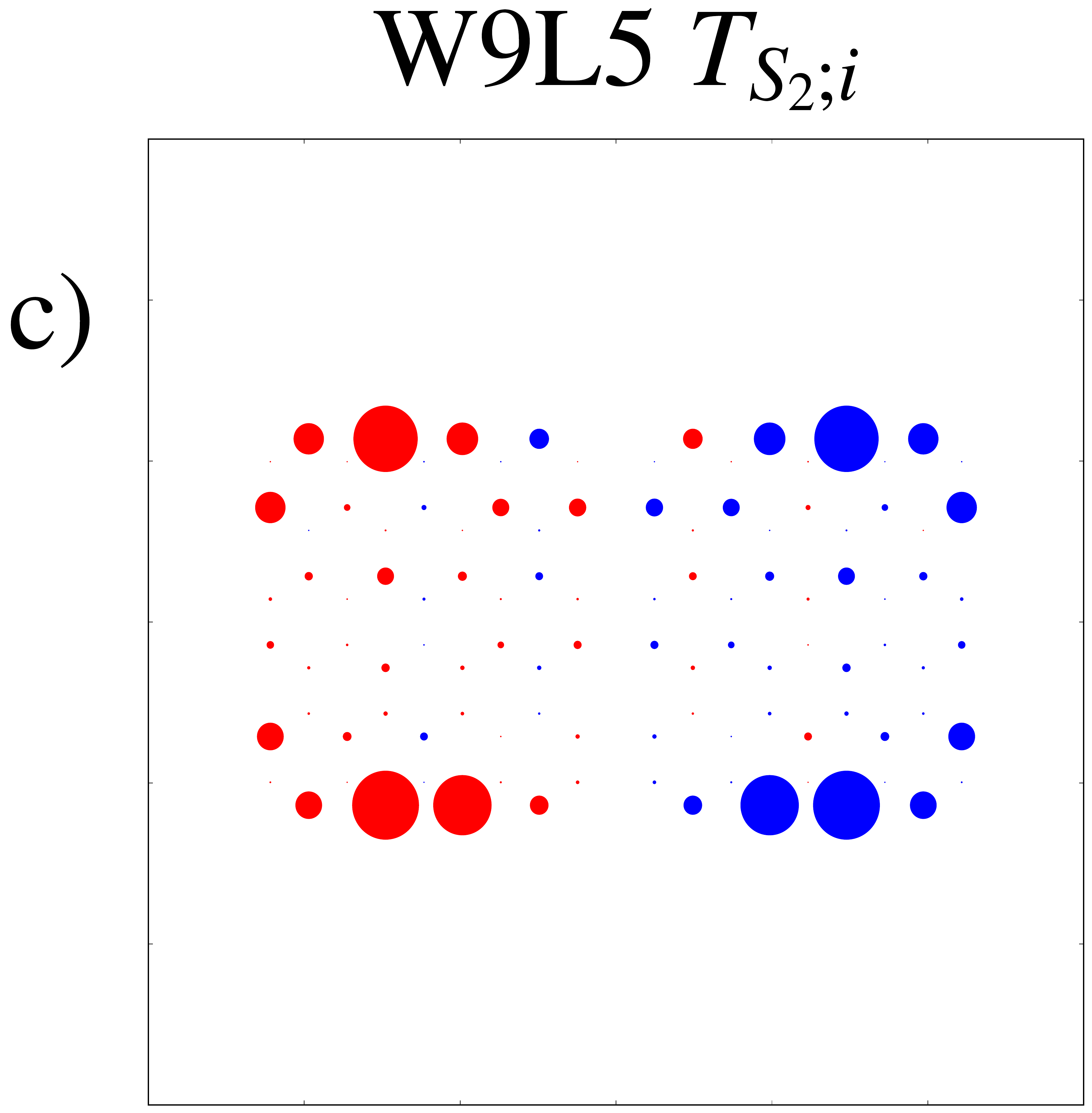}
\end{tabular}
\begin{tabular}{rr}
\includegraphics[width=0.19\textwidth]{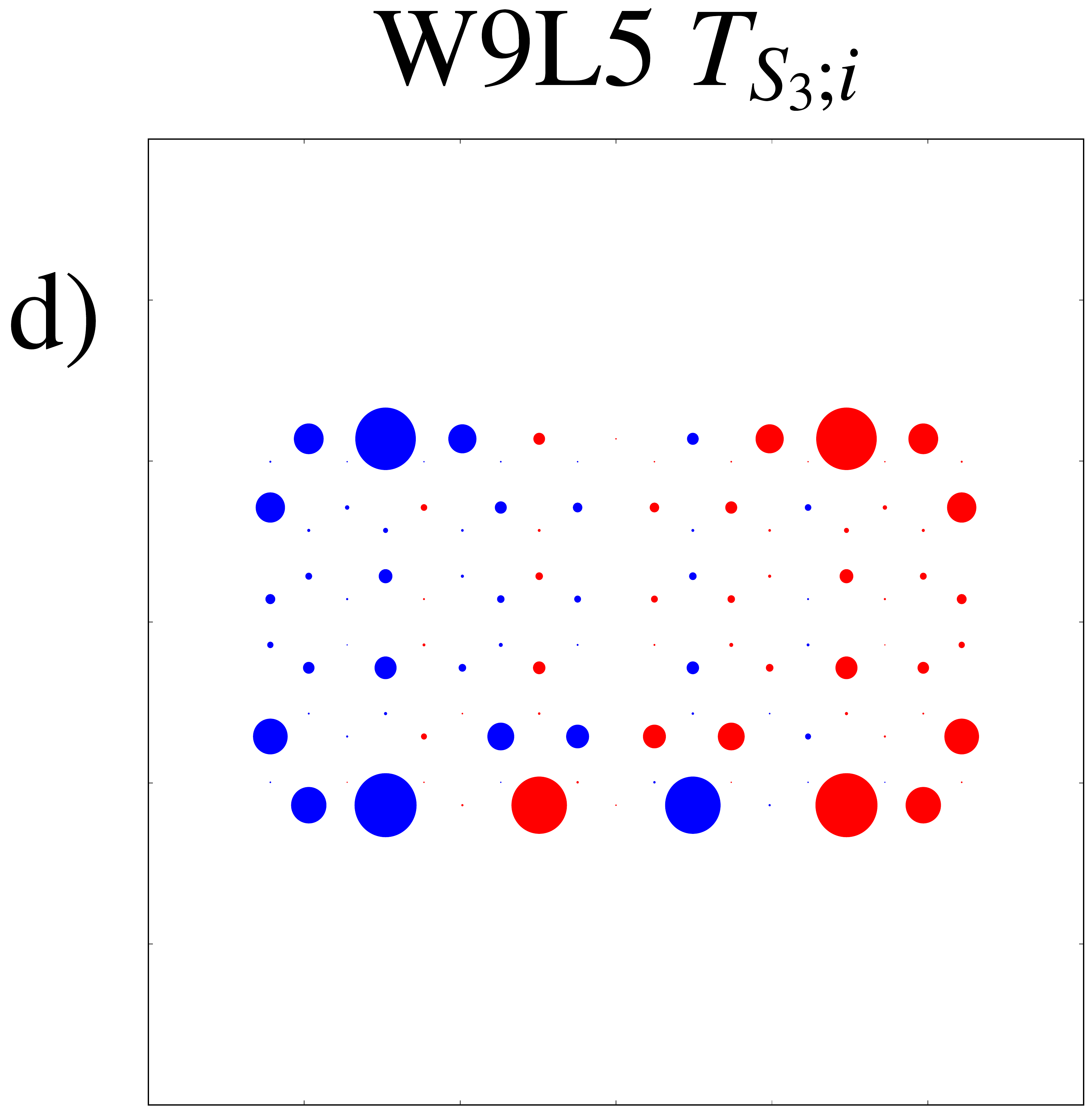} &
\includegraphics[width=0.19\textwidth]{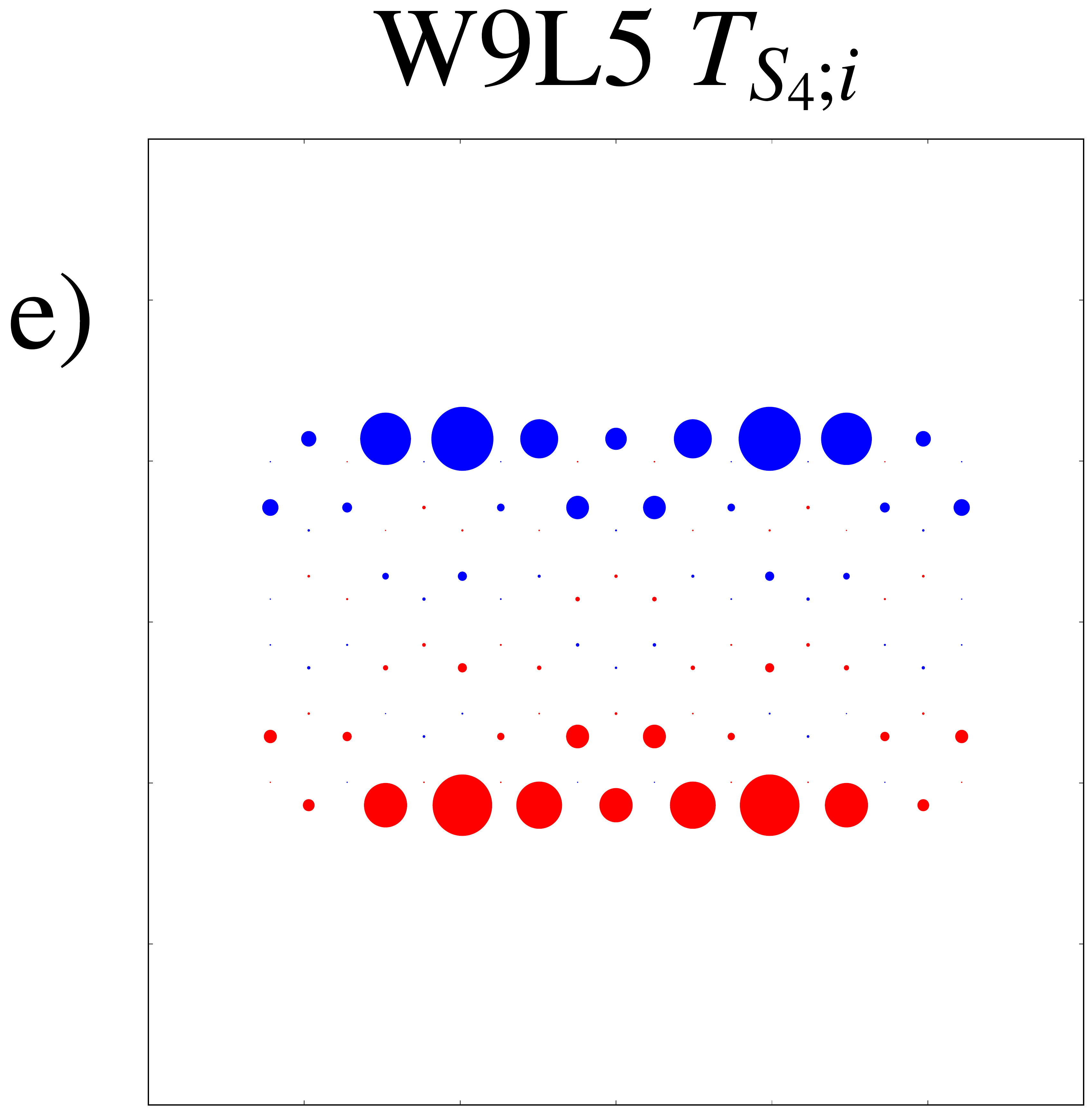}
\end{tabular}
\caption{Plot of a) the profile of an impurity placed in the zig-zag bottom site of the W9L5 flake with the same parameter as Fig. \ref{fig9}. Plot of b) $T_{S_{1};i}$, c) $T_{S_{2};i}$, d) $T_{S_{3};i}$, and e) $T_{S_{4};i}$ for the W9L5 flake with an impurity potential from a). The charge distributions are still concentrated on the zig-zag edges for these transitions.}
\label{fig10}
\end{figure*}

\begin{figure}[htb!]
\begin{center}
\includegraphics[width=1.0\columnwidth]{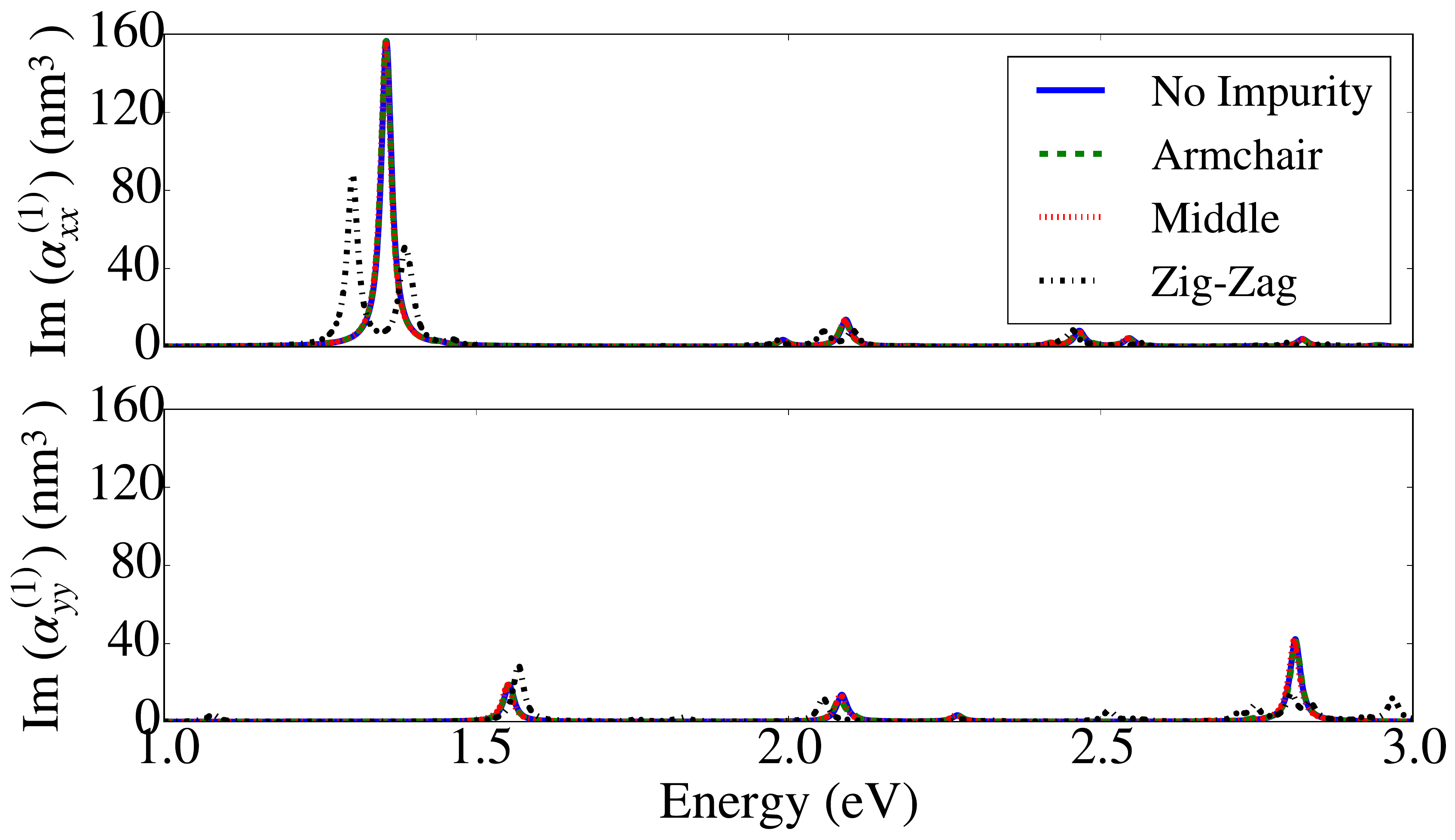}
\caption{Absorption spectrum of the W9L5 flake with varied impurity locations and a weaker strength. The parameters used to model the impurity potential were $\overline{\varepsilon}_{max} = t/5$ and $\tau = l_{b}$.}
\label{fig9b}
\end{center}
\end{figure}

In order to determine the robustness of the influence of impurities on the zig-zag edges of a GF on their optical properties, we consider a shorter range impurity potential, as well as a weaker one.
In Fig. \ref{fig5a} we plot the absorption spectrum for the W3L11 flake with a zig-zag impurity of range $\tau = l_{b}/5.0$, which corresponds to an impurity potential essentially confined to a single site. 
It shows that even a shorter range impurity potential centered on the zig-zag edge leads to a significant change in the absorption spectrum, as well as in the joint density of states, and they also mix otherwise dark transitions with bright HF excitations. 
In Fig. \ref{fig5a} we plot the absorption spectrum for the W3L11 flake with a zig-zag impurity of reduced strength $\overline{\varepsilon}_{max} = t/5$, while setting the range of influence to $\tau = l_{b}$. 
It shows that even a weaker impurity on the zig-zag edge has a significant effect on the absorption spectrum of rectangular GFs.

%\begin{figure*}
%\begin{center}
%
%\caption{Absorption spectrum of the W11L3 flake with various local impurities; the absorption remains relatively unchanged for impurities located in the middle of the flake, but placing an impurity on one of the zig-zag edges results in several new peaks. The zig-zag impurity also induces a very weak, low energy absorption with a transition dipole moment oriented in the $\hat{\mbf{y}}$ direction, the short axis of the flake. Even a very light impurity on the zig-zag edge can have a dramatic impact on the absorption spectrum of these flakes.  The parameters used to model the impurity were $\overline{\varepsilon}_{max} = t/5$ and $\tau = l_{b}$.}
%\label{fig7a}
%\end{center}
%\end{figure*}

We now analyze a GF with larger zig-zag edges than armchair edges, namely the W11L3 flake.
In Fig. \ref{fig7} we plot the absorption spectrum for the W11L3 flake with impurities with potential strength $\overline{\varepsilon}_{max} = t/3$ and range $\tau = l_{b}$. 
As we found for the W3L11 flake, an impurity located in the middle of the W11L3 flake, or on its armchair edges, has almost no effect on the absorption spectrum unless the impurity potential extends to the zig-zag edges. Impurities located on the zig-zag edges, however, can have a significant effect on the absorption spectrum of these flakes. This is because, like the W3L11 flake, the charge distributions involved in the low lying transitions are concentrated on the zig-zag edges of the flake.
Impurities on the zig-zag edges lead to a splitting of the first absorption peak into three smaller ones, and produces a very weak low energy absorption peak corresponding to a transition whose transition dipole moment is polarized along the short (here $\hat{\mbf{y}}$) axis of the flake. 
This weak low energy absorption peak is due to a dark transition that becomes mixed with a bright HF single excitation involving the edge modes in the presence of the impurity. 
In Fig. \ref{fig8} we show the quantities $T_{S_{1};i}$, $T_{S_{2};i}$, $T_{S_{3};i}$, and $T_{S_{4};i}$ for the W11L3 flake with a zig-zag impurity.
Much like the pristine GFs, the spatial profiles of these transitions are concentrated on the zig-zag edges, but here with reduced electron concentration around the center of the impurity potential. 
%
%In the pristine flake, the transition dipoles are oriented along one of the main axes, but the armchair impurities lead to transition dipole moments slightly turned from the main axes, which is due to $\text{Im} (\alpha^{(1)}_{ij})$ being non-zero for $i\neq j$. 
%
Impurities away from the zig-zag edges but with an extended impurity potential that reaches those edges can also impact the optical properties of GFs.
In Fig. \ref{fig7a}, we show the dependence of the absorption spectrum of the W11L3 flake on the range of the impurity potential as well as strength of the impurity potential. 
A shorter range impurity potential ($\tau = l_{b}/5$) on an armchair edge does not change the absorption spectrum, nor does it have a significant effect on the joint density of states, while a shorter range impurity potential on the zig-zag flake does. 
Even a weaker strength impurity potential ($\overline{\varepsilon}_{max} = t/5$) on the zig-zag edge leads to a significant change in the absorption spectrum and the joint density of states. 

%\begin{figure*}
%\begin{center}
%\includegraphics[scale=0.4]{gfW9L5_loc_dis_abs.pdf}
%\caption{Absorption spectrum of the W9L5 flake with various local impurity potentials; the absorption remains relatively unchanged for impurities located on either the armchair edges or in the middle of the flake, but placing an impurity on one of the zig-zag edges results in several new peaks, including a relatively weak low energy absorption which has a dipole moment that is polarized in the $\hat{\mbf{y}}$ direction.}
%\label{fig9a}
%\end{center}
%\end{figure*}

Lastly we analyze the effects of impurities on the larger W9L5 flake. 
In Fig. \ref{fig9}, we plot the absorption spectrum for a impurity potential with strength $\overline{\varepsilon}_{max} = t/3$, and range $\tau = l_{b}$, located at different locations on the flake. 
Again, an impurity on the zig-zag edge has a significant effect on the absorption spectrum, while impurities on the armchair edges, or the middle of the flake, have almost no impact. This is because the charge distributions involved in the low lying transitions of the system are concentrated mainly on the zig-zag edge.
The zig-zag impurity leads to several peaks within the proximity of the lowest energy absorption peak of the pristine GF. 
Similarly to the W$n$L3 family with zig-zag impurities, there is a very weak low energy absorption peak associated with a transition whose transition dipole moment is polarized along the short (here $\hat{\mbf{y}}$) axis of the flake. 
This peak is associated with a dark transition that becomes mixed with several bright HF excitations, including transitions between edge modes. 
%Within the proximity of the first absorption peak, the impurity potential leads to the mixing of previously dark states with bright HF excitations, and leads to the splitting of an initially bright state into two different bright states. 
The impurity potential also significantly enhances (by a factor of 2.5) the previously weak absorption peak with an associated transition dipole moment polarized along the short (here $\hat{\mbf{y}}$) axis of the flake by mixing the excited state with several other bright HF excitations. 
We plot the transition densities $T_{S_{1};i}$, $T_{S_{2};i}$, $T_{S_{3};i}$, and $T_{S_{4};i}$ in Fig. \ref{fig10}, which shows they are still concentrated on the zig-zag edges. In Fig. \ref{fig9b}, we explore the dependence of the absorption spectrum on the strength of the impurity potential.
Even a weaker impurity potential placed on the zig-zag edge can have a significant impact on the absorption spectrum.

Ultimately, impurities on the zig-zag edge have a significant effect on the low energy absorption spectrum of the GFs studied in this paper. This is because the charge distributions involved in the low energy transitions are concentrated on the zig-zag edges. Impurities located in the middle of the flake, or on the armchair edges, have essentially no impact as they do not affect the charge distributions involved in the low energy transitions in these systems. Even though the impurity potential can often lead to the shift of the energies of peaks, and can turn dark transitions bright, it can sometimes enhance certain absorption features, making the role of impurities potentially beneficial to certain applications.

\section{\label{sec-conc} Conclusion}

We have calculated the optical properties of several rectangular graphene flakes (GFs), taking into account electron correlations beyond the mean-field level. Including these correlations is essential to accurately describe the low energy absorption of these flakes, as mean-field theory alone cannot accurately predict their optical properties. 
We find that the first absorption peak invariably corresponds to a transition dipole moment polarized along the longest axis of the flake. 
We also find that the electron concentration for the low energy transitions are always concentrated on the zig-zag edges, regardless of whether or not the zig-zag edges are longer than the armchair edges. 
The zig-zag edges of rectangular GFs thus play a primary role in the optical absorption of the flakes. 
%
%We also find that for GFs with longer zig-zag edges, the lowest energy transition is associated with a state whose dominant contribution is the removal of an electron from the HOHF state to the LUHF one. 
%In contrast, for GFs with longer armchair edges, the lowest energy absorption peak is associated with a state which is a superposition of several HF double excitations. 
% This result can only be analyzed using the CI calculations we used as it cannot be reproduced by a mean field theory approach. 

We also investigated the effect of impurities on the optical properties of rectangular GFs by placing impurities potentials of different spatial ranges and strengths on different locations.
We find that the effect of impurities on the optical properties of these GFs strongly depends on the location of the impurity potential. 
Impurities on the zig-zag edges have a significant impact on the optical properties of these GFs, while impurities on the ``bulk'' region or their armchair edges have a negligible impact on the frequencies and the nature of the optical transitions. 
% The optical properties of the GFs studied in this paper are robust to the presence of impurity potentials centered on the armchair edges or in the middle of the flake. 
% The impurity potentials centered on the zig-zag can often enhance, or suppress particular absorption features; these impurities can often turn dark states bright. 
% Even very weak impurity potentials, or those located only on the zig-zag edge, can have a dramatic impact on the absorption spectrum of these GFs.
We expect that understanding these qualitative features will be central in the design of any GF devices, both when it is desirable to avoid the effects of impurities, and when it is desirable to exploit their effect on the optical properties of the GFs.

\appendix

\section{\label{app-Ham} Full Hamiltonian in the electron/hole basis}

In this appendix, we rewrite the total Hamiltonian in the electron-hole basis, which is an important step in our configuration interaction calculations (\ref{CI_gr_st}).
 
\subsection{The tight-binding and impurity Hamiltonian in the electron/hole basis}

The tight-binding Hamiltonian (\ref{TBeq}), combined with the impurity potential (\ref{impham}), can be written as
\beq
H_{TB} + H_{imp} =  - t \sum_{\la i,j \ra, \sigma} c^{\dg}_{i\sigma} c_{j\sigma} + \sum_{i\sigma} \overline{\varepsilon}_{i} c^{\dg}_{i\sigma} c_{i\sigma}.
\eeq
Moving to the basis defined in (\ref{eq-sphf},\ref{eq-sphf2}), this Hamiltonian can be written as 
\beq
H_{TB} + H_{imp} = \sum_{mm'\sigma} \tilde{\kappa}_{mm'} C^{\dg}_{m\sigma} C_{m'\sigma}  \label{tbhf},
\eeq
where
\beq
\tilde{\kappa}_{mm'} =   \sum_{i}\overline{\varepsilon}_{i}M_{m\sigma,i}M^{*}_{m'\sigma,i}  -t \sum_{\la i,j \ra} M_{m\sigma,i} M^{*}_{m'\sigma,j}.
\eeq
Then, we can rewrite (\ref{tbhf}) in the electron-hole basis (\ref{hfhole}), as
\begin{align}
H_{TB} + H_{imp} = & \sum_{mm'\sigma} \tilde{\kappa}_{mm'} a^{\dg}_{m\sigma} a_{m'\sigma}  - \sum_{mm'\sigma} \tilde{\kappa}_{mm'}  b^{\dg}_{m'\sigma} b_{m\sigma} \nonumber \\
& + \sum_{mm'\sigma} \tilde{\kappa}_{mm'} \left( a^{\dg}_{m\sigma} b^{\dg}_{m'\tilde{\sigma}} + b_{m\tilde{\sigma}} a_{m'\sigma} \right) + \sum_{m\sigma} \tilde{\kappa}_{mm\sigma}.
\end{align}

\subsection{The Hubbard Hamiltonian in the electron/hole basis}

The Hubbard Hamiltonian (\ref{hubb}) can be written as
\beq
H_{Hu} = U \sum_{i} n_{i\ua} n_{i\da}.
\eeq
In the basis defined in (\ref{eq-sphf},\ref{eq-sphf2}), we can write this as
\beq
H_{Hu} = \sum_{mm'pp'} \Gamma_{mm'pp'} C^{\dg}_{m\ua} C_{m'\ua} C^{\dg}_{p\da} C_{p'\da},
\eeq
where
\beq
\Gamma_{mm'pp'} = U\sum_{i}M_{m\ua i} M^{*}_{m' \ua i} M_{p \da, i} M^{*}_{p'\da,i}.
\eeq
Moving to an electron-hole basis, the Hubbard Hamiltonian can be written as
\beq
H_{Hu} = H_{Hu;0} + H_{Hu;1} + H_{Hu;2} + H_{Hu;3} + H_{Hu;4}.
\eeq
The first term can be written as
\beq
H_{Hu;0} = \sum_{\substack{m \in \text{ filled} \\ n \in \text{ filled}}} \Gamma_{mmnn}.
\label{eq-hu0}
\eeq
Eq. (\ref{eq-hu0}) accounts for the Coulomb repulsion of the nominal vacuum. 

\begin{widetext}

The second part of the Hamiltonian is
\begin{align}
H_{Hu;1} = & \sum_{\substack{mm' \\ p \in \text{ filled}}} \Gamma_{mm'pp} a^{\dg}_{m\ua}a_{m'\ua} + \sum_{\substack{pp' \\ m \in \text{ filled}}} \Gamma_{mmpp'} a^{\dg}_{p\da}a_{p'\da}  - \sum_{\substack{pp' \\ m \in \text{ filled}}} \Gamma_{mmpp'} b^{\dg}_{p'\ua} b_{p\ua} - \sum_{\substack{mm' \\ p \in \text{ filled}}} \Gamma_{mm'pp} b^{\dg}_{m'\da} b_{m\da} \nonumber \\
&  + \sum_{\substack{mm' \\ p \in \text{ filled}}} \Gamma_{mm'pp} a^{\dg}_{m\ua} b^{\dg}_{m'\da} + \sum_{\substack{pp' \\ m \in \text{ filled}}} \Gamma_{mmpp'} a^{\dg}_{p\da} b^{\dg}_{p'\ua} +  \sum_{\substack{mm' \\ p \in \text{ filled}}} \Gamma_{mm'pp} b_{m\da} a_{m'\ua} + \sum_{\substack{pp' \\ m \in \text{ filled}}} \Gamma_{mmpp'} b_{p\ua} a_{p'\da}.  \label{eq-hu1}
\end{align}
Eq. \ref{eq-hu1} contains the single particle terms that play a role in the matrix elements of both single and double excitations. The third part of Hubbard Hamiltonian is
\begin{align}
H_{Hu;2} = & \sum_{mm'pp'} \Gamma_{mm'pp'} \left(a^{\dg}_{p\da} b_{m\da} - a^{\dg}_{m\ua} b_{p\ua} \right) a_{m'\ua} a_{p'\da} + \sum_{mm'pp'} \Gamma_{mm'pp'} a^{\dg}_{m\ua} a^{\dg}_{p\da} \left( b^{\dg}_{p'\ua} a_{m'\ua} - b^{\dg}_{m'\da} a_{p'\da} \right) \nonumber \\
& + \sum_{mm'pp'} \Gamma_{mm'pp} \left(b^{\dg}_{m'\da}a_{p'\da} - b^{\dg}_{p'\ua} a_{m'\ua}  \right) b_{p\ua} b_{m\da} + \sum_{mm'pp'} \Gamma_{mm'pp} b^{\dg}_{m'\da} b^{\dg}_{p'\ua} \left( a^{\dg}_{p\da} b_{m\da} - a^{\dg}_{m\ua} b_{p\ua} \right). \label{eq-hu2}
\end{align}
\end{widetext}
Eq. \ref{eq-hu2} has matrix elements between single and double excitations. The fourth part of the Hubbard Hamiltonian is
\begin{align}
 H_{Hu;3} = & - \sum_{mm'pp'} \Gamma_{mm'pp'} \left( a^{\dg}_{p\da} b^{\dg}_{m'\da} b_{m\da} a_{p'\da} + a^{\dg}_{m\ua} b^{\dg}_{p'\ua} b_{p \ua} a_{m'\ua}\right) \nonumber \\
& + \sum_{mm'pp'} \Gamma_{mm'pp'} \left(a^{\dg}_{p\da} b^{\dg}_{p'\ua}  b_{m\da} a_{m'\ua} +   a^{\dg}_{m\ua} b^{\dg}_{m'\da}  b_{p\ua} a_{p'\da} \right) \nonumber \\
& - \sum_{mm'pp'} \Gamma_{mm'pp'} \left(b_{m\da} b_{p\ua} a_{m'\ua} a_{p'\da} +  a^{\dg}_{m\ua} a^{\dg}_{p\da} b^{\dg}_{m'\da} b^{\dg}_{p'\ua} \right). 
\label{eq-hu3}
\end{align}
Eq. (\ref{eq-hu3}) is the part of the Hamiltonian that has a contribution to the matrix elements between single excitations, between the ground state and double excitations, as well as between different double excitations. The last part of the Hubbard Hamiltonian is
\begin{align}
H_{Hu;4} = & \sum_{mm'pp'} \Gamma_{mm'pp'} a^{\dg}_{p\da} a^{\dg}_{m\ua} a_{m'\ua}  a_{p'\da} \nonumber \\
 & + \sum_{mm'pp'} \Gamma_{mm'pp'} b^{\dg}_{m'\da} b^{\dg}_{p'\ua} b_{p\ua} b_{m\da}. 
\label{eq-hu4}
\end{align}
The term (\ref{eq-hu4}) has matrix elements between double excitations only. 

\subsection{The extended Hubbard Hamiltonian in the electron/hole basis}

The extended Hubbard Hamiltonian can be written as
\beq
H_{ext} = H_{ee} + H_{en} + H_{nn},
\eeq
where
\begin{align}
H_{ee} = & \frac{1}{2} \sum_{\substack{i\neq j \\ \sigma \sigma'}} V_{ij} n_{i\sigma} n_{j\sigma'}, \label{eq-hee} \\
 H_{en} = & -\frac{1}{2} \sum_{\substack{i \neq j \\ \sigma}} V_{ij} \left( n_{i\sigma} + n_{j\sigma} \right), \label{eq-hen} \\
 H_{nn} = & \frac{1}{2} \sum_{i\neq j} V_{ij}. \label{eq-hnn}
\end{align}
The term (\ref{eq-hee}) describes the long-range interaction between the electrons, (\ref{eq-hen}) describes the electron-nuclei interaction, and the term (\ref{eq-hnn}) describes the nuclei-nuclei interaction, which in our model is a constant. 
\subsubsection{$H_{ee}$ in the Electron/Hole Basis}
The extended Hubbard electron-electron repulsion Hamiltonian is
\beq
H_{ee} = \frac{1}{2}\sum_{i\neq j \sigma \sigma'}V_{ij} n_{i\sigma} n_{j\sigma'}.
\eeq
Rewriting this in the HF basis,
\beq
H_{ee} = \sum_{\substack{mm'nn' \\ \sigma \sigma'}}\tilde{\Phi}_{mm'nn'} C^{\dg}_{m\sigma} C_{m'\sigma} C^{\dg}_{n\sigma'} C_{n'\sigma'} \label{HFEXT},
\eeq
where
\beq
\Phi_{mm'nn'} = \frac{1}{2} \sum_{i\neq j} V_{ij} M_{m\sigma,i} M^{*}_{m'\sigma,i} M_{n\sigma',j} M^{*}_{n'\sigma',j}.
\eeq
Moving to the electron-hole basis and normal ordering, the Hamiltonian can be written as
\beq
H_{ee} = H_{ee;0} + H_{ee;1} + H_{ee;2} + H_{ee;3} + H_{ee;4}.
\eeq
\begin{widetext}
The first part of the Hamiltonian can be written as
\beq
H_{ee;0} =  \sum_{\substack{\sigma \sigma' \\ m,n \in \text{ filled}}} \Phi_{mmnn} + \sum_{\substack{m \in \text{ filled} \\ n \in \text{ unfilled} \\ \sigma}} \Phi_{mnnm}. \label{hee0}
\eeq
The term $H_{ee;0}$ (\ref{hee0}) represents the long-range Coulomb repulsion of the initial ground state. The second part of the Hamiltonian is

\begin{align}
H_{ee;1} = & \sum_{\substack{mn' \sigma \\ n \text{ $\in$ unfilled}}} \Phi_{mnnn'} a^{\dg}_{m\sigma} a_{n'\sigma}  - \sum_{\substack{nn' \sigma \sigma' \\ m \in \text{ filled}}}\Phi_{mmnn'} b^{\dg}_{n'\sigma}b_{n\sigma} - \sum_{\substack{mm'\sigma \sigma' \\ n \in \text{ filled}}} \Phi_{mm'nn} b^{\dg}_{m'\sigma}b_{m\sigma}  
 + \sum_{\substack{m'n \sigma \\ m \in \text{ filled}}} \Phi_{mm'nm} b^{\dg}_{m'\sigma} b_{n\sigma} 
\nonumber \\
&  + \sum_{\substack{mn'\sigma \\ n \text{ $\in$ unfilled}}} \Phi_{mnnn'} a^{\dg}_{m\sigma}b^{\dg}_{n'\tilde{\sigma}'} + \sum_{\substack{mn'\sigma \\ n \in \text{ unfilled}}} \Phi_{mnnn'} b_{m\tilde{\sigma}}a_{n'\sigma}  +  \sum_{\substack{mm'\sigma\sigma' \\ n \in \text{ filled}}} \Phi_{mm'nn} a^{\dg}_{m\sigma}a_{m'\sigma}  - \sum_{\substack{mn' \sigma\\ n \in \text{ unfilled}}} \Phi_{mnnn'} b^{\dg}_{n'\sigma} b_{m\sigma} 
\nonumber \\
& +  \sum_{\substack{nn' \sigma \sigma' \\ m \in \text{ filled}}}\Phi_{mmnn'} a^{\dg}_{n\sigma}a_{n'\sigma} - \sum_{\substack{m'n \sigma \\ m \in \text{ filled}}} \Phi_{mm'nm} a^{\dg}_{n\sigma} a_{m'\sigma}
 + \sum_{\substack{mm' \sigma \sigma' \\ n \in \text{ filled}}} \Phi_{mm'nn} a^{\dg}_{m\sigma}b^{\dg}_{m'\tilde{\sigma}}
 +  \sum_{\substack{mm' \sigma \sigma' \\ n \in \text{ filled}}}\Phi_{mm'nn}b_{m\tilde{\sigma}}a_{m'\sigma} 
\nonumber \\
& + \sum_{\substack{m'n \sigma \\ m \in \text{ filled}}} \Phi_{mm'nm}a_{m'\sigma}b_{n\tilde{\sigma}} + \sum_{\substack{ nn' \sigma \sigma' \\ m \in \text{ filled}}} \Phi_{mmnn'}a^{\dg}_{n\sigma'} b^{\dg}_{n'\tilde{\sigma}'} + \sum_{\substack{ m'n \sigma \\ m \in \text{ filled}}}\Phi_{mm'nm} b^{\dg}_{m'\tilde{\sigma}}a^{\dg}_{n\sigma}
 + \sum_{\substack{nn' \sigma \sigma' \\ m \in \text{ filled} }} \Phi_{mmnn'} b_{n\tilde{\sigma}}a_{n'\sigma}. \label{Hee1}
\end{align}
The term $H_{ee;1}$ (\ref{Hee1}) represents the single-particle terms that play a role in the matrix elements of both single and double excitations. The third part of the Hamiltonian is
\begin{align}
H_{ee;2} = & \sum_{mm'nn'\sigma\sigma'} \Phi_{mm'nn'} a^{\dg}_{m\sigma}a^{\dg}_{n\sigma'}b^{\dg}_{n'\tilde{\sigma}'}a_{m'\sigma} + \sum_{mm'nn'\sigma\sigma'} \Phi_{mm'nn'} a^{\dg}_{m\sigma} a_{n'\sigma'} a_{m'\sigma}b_{n\tilde{\sigma}'} 
 + \sum_{mm'nn'\sigma\sigma'} \Phi_{mm'nn'} a^{\dg}_{n\sigma'} a^{\dg}_{m\sigma} b^{\dg}_{m'\tilde{\sigma}} a_{n'\sigma'} 
\nonumber \\
& +  \sum_{mm'nn'\sigma\sigma'} \Phi_{mm'nn'} a^{\dg}_{n\sigma'} b_{m\tilde{\sigma}}a_{m'\sigma}a_{n'\sigma'} 
 + \sum_{mm'nn'\sigma\sigma'} \Phi_{mm'nn'} a^{\dg}_{m\sigma}b^{\dg}_{n'\sigma'}b^{\dg}_{m'\tilde{\sigma}}b_{n\sigma'} + \sum_{mm'nn'\sigma\sigma'} \Phi_{mm'nn'} b^{\dg}_{n'\sigma'}a_{m'\sigma}b_{m\tilde{\sigma}}b_{n\sigma'} 
\nonumber \\
& + \sum_{mm'nn'\sigma\sigma'} \Phi_{mm'nn'} b^{\dg}_{m'\sigma}b^{\dg}_{n'\tilde{\sigma}'}a^{\dg}_{n\sigma'}b_{m\sigma} - \sum_{mm'nn'\sigma\sigma'} \Phi_{mm'nn'}b^{\dg}_{m'\sigma}b_{m\sigma}b_{n\tilde{\sigma}'}a_{n'\sigma'}. 
\label{Hee2}
\end{align}
The term $H_{ee;2}$ (\ref{Hee2}) can have non-zero matrix elements between single and double excitations. The fourth part of the Hamiltonian is
\begin{align}
H_{ee;3} = & - \sum_{mm'nn'\sigma\sigma'} \Phi_{mm'nn'} a^{\dg}_{m\sigma}b^{\dg}_{n'\sigma'}b_{n\sigma'}a_{m'\sigma}  - \sum_{mm'nn'\sigma\sigma'} \Phi_{mm'nn'} a^{\dg}_{n\sigma'}b^{\dg}_{m'\sigma}b_{m\sigma} a_{n'\sigma'}   + \sum_{mm'nn'\sigma\sigma'} \Phi_{mm'nn'} a^{\dg}_{m\sigma} b^{\dg}_{m'\tilde{\sigma}} b_{n\tilde{\sigma}'} a_{n'\sigma'} \nonumber \\
& + \sum_{mm'nn'\sigma\sigma'} \Phi_{mm'nn'}  a^{\dg}_{n\sigma'}b^{\dg}_{n'\tilde{\sigma}'} b_{m\tilde{\sigma}} a_{m'\sigma}   + \sum_{mm'nn'\sigma\sigma'} \Phi_{mm'nn'} b_{m\tilde{\sigma}} a_{m'\sigma} b_{n\tilde{\sigma}'} a_{n'\sigma'} + \sum_{mm'nn'\sigma\sigma'} \Phi_{mm'nn'} a^{\dg}_{m\sigma} b^{\dg}_{m'\tilde{\sigma}} a^{\dg}_{n\sigma'} b^{\dg}_{n'\tilde{\sigma}'}. 
\label{Hee3}
\end{align}
\end{widetext}
The term $H_{ee;3}$ (\ref{Hee3}) contributes to the matrix elements between single excitations, between the ground state and double excitations, as well as between different double excitations. The fifth part of the Hamiltonian is
\begin{align}
H_{ee;4} = & \sum_{mm'nn'\sigma \sigma'} \Phi_{mm'nn'} a^{\dg}_{n\sigma'} a^{\dg}_{m\sigma} a_{m'\sigma} a_{n'\sigma'} \nonumber \\
& + \sum_{mm'nn'\sigma \sigma'} \Phi_{mm'nn'} b^{\dg}_{n'\sigma'} b^{\dg}_{m'\sigma} b_{m\sigma} b_{n\sigma'}. 
\label{Hee4}
\end{align}
The term $H_{ee;4}$ (\ref{Hee4}) contributes to the matrix elements between double excitations only. 

\subsubsection{$H_{en}$ in the Electron/Hole Basis}

The electron-nuclei Hamiltonian is given by
\beq
H_{en} = -\frac{1}{2} \sum_{i \neq j, \sigma} V_{ij} \left( n_{i\sigma} + n_{j\sigma} \right).
\eeq
In the basis defined in (\ref{eq-sphf},\ref{eq-sphf2}), this is
\beq
H_{en} = \sum_{mm'\sigma} \tilde{\phi}_{mm'} C^{\dg}_{m\sigma} C_{m'\sigma} \label{HFEN},
\eeq
where
\beq
\overline{\phi}_{mm'} = - \frac{1}{2}\sum_{i\neq j} V_{ij} \left( M_{m\sigma,i} M_{m'\sigma,j} + M_{m\sigma,j} M^{*}_{m'\sigma,j} \right).
\eeq
Moving to the electron-hole basis, we can write (\ref{HFEN}) as
\begin{align}
H_{en} = &  \sum_{\substack{m \in \text{ filled},  \sigma}} \overline{\phi}_{mm}
+ \sum_{mm'\sigma} \overline{\phi}_{mm'} a^{\dg}_{m\sigma} a_{m'\sigma}  + \sum_{mm'\sigma} \overline{\phi}_{mm'} a^{\dg}_{m\sigma} b^{\dg}_{m'\tilde{\sigma}} \nonumber \\
& + \sum_{mm'\sigma} \overline{\phi}_{mm'} b_{m\tilde{\sigma}}a_{m'\sigma}   -\sum_{mm'\sigma} \overline{\phi}_{mm'} b^{\dg}_{m'\sigma} b_{m\sigma }. 
\label{Hen}
\end{align}
The electron-nuclei interaction (\ref{Hen}) contributes to matrix elements between single excitations, matrix elements between double excitations, and matrix elements between single and double excitations and single excitations and the nominal vacuum. 

%\newpage

%\bibliographystyle{unsrt}
%\bibliographystyle{apsrev4-1}

\bibliography{gf_imp}

\end{document}